\documentclass[aps,pra,reprint,superscriptaddress]{revtex4-1}
\usepackage{natbib}

\usepackage{color}
\usepackage{amsmath}
\usepackage{amssymb}
\usepackage{framed}
\definecolor{shadecolor}{rgb}{0.85, .85, 0.85}
\usepackage{graphicx}

\definecolor{darkgreen}{rgb}{0.05, .65, 0.05}

\def\dlangle{\langle\!\langle}
\def\drangle{\rangle\!\rangle}
\def\bigdlangle{{\big\langle\kern-.23em\big\langle}}
\def\bigdrangle{{\big\rangle\kern-.23em\big\rangle}}
\def\Bigdlangle{{\Big\langle\kern-.35em\Big\langle}}
\def\Bigdrangle{{\Big\rangle\kern-.35em\Big\rangle}}
\def\biggdlangle{{\bigg\langle\kern-.35em\bigg\langle}}
\def\biggdrangle{{\bigg\rangle\kern-.35em\bigg\rangle}}
\def\Biggdlangle{{\Bigg\langle\kern-.35em\Bigg\langle}}
\def\Biggdrangle{{\Bigg\rangle\kern-.35em\Bigg\rangle}}
\def\dexpct#1{\dlangle{#1}\drangle}
\def\bigdexpct#1{\bigdlangle{#1}\bigdrangle}
\def\Bigdexpct#1{\Bigdlangle{#1}\Bigdrangle}

\def\ds{\displaystyle}
\def\intinf{\int_{-\infty}^\infty\!\!}
\def\intzinf{\int_{0}^\infty\!\!}

\def\linklangle{{\langle\kern-.28em\scalebox{0.35}[1]{$-$}\kern-.18em\langle}}
\def\linkrangle{{\rangle\kern-.18em\scalebox{0.35}[1]{$-$}\kern-.28em\rangle}}
\def\biglinklangle{{\kern.23em\big\langle\kern-.59em\scalebox{0.35}[1]{$-$}\kern-.38em\big\langle\kern.23em}}
\def\biglinkrangle{{\kern.23em\big\rangle\kern-.38em\scalebox{0.35}[1]{$-$}\kern-.59em\big\rangle\kern.23em}}
\def\Biglinklangle{{\kern.0em\Big\langle\kern-.49em\scalebox{0.38}[1]{$-$}\kern-.18em\Big\langle\kern.0em}}
\def\Biglinkrangle{{\kern.0em\Big\rangle\kern-.18em\scalebox{0.38}[1]{$-$}\kern-.49em\Big\rangle\kern.0em}}
\def\bigglinklangle{{\kern.0em\bigg\langle\kern-.62em\scalebox{0.54}[1]{$-$}\kern-.18em\bigg\langle\kern.0em}}
\def\bigglinkrangle{{\kern.0em\bigg\rangle\kern-.18em\scalebox{0.54}[1]{$-$}\kern-.62em\bigg\rangle\kern.0em}}
\def\Bigglinklangle{{\kern.0em\Bigg\langle\kern-.66em\scalebox{0.66}[1]{$-$}\kern-.22em\Bigg\langle\kern.0em}}
\def\Bigglinkrangle{{\kern.0em\Bigg\rangle\kern-.22em\scalebox{0.66}[1]{$-$}\kern-.66em\Bigg\rangle\kern.0em}}

\def\tr{\ensuremath{\mathrm{tr}}}

\def\vect#1{\ensuremath{\mathbf{#1}}}
\def\op#1{\ensuremath{\hat{#1}}}
\def\sgn{\ensuremath{\mathrm{sgn}}}

\def\cT{\ensuremath{\mathcal{T}}}

\def\rA{\ensuremath{\mathbf{r}_\mathrm{\scriptscriptstyle A}}}
\def\x0{\ensuremath{\mathbf{x}_\mathrm{0}}}
\def\epsr{\ensuremath{\epsilon_{\mathrm{r}}}}
\def\mur{\ensuremath{\mu_{\mathrm{r}}}}
\def\subEM{_\mathrm{\scriptscriptstyle EM}}
\def\subTE{_\mathrm{\scriptscriptstyle TE}}
\def\subTM{_\mathrm{\scriptscriptstyle TM}}
\def\subCP{_\mathrm{\scriptscriptstyle CP}}
\def\supTE{^\mathrm{\scriptscriptstyle (TE)}}
\def\supTM{^\mathrm{\scriptscriptstyle (TM)}}
\def\ZTE{\ensuremath{Z\subTE}}
\def\VTE{\ensuremath{V\subTE}}
\def\ETE{\ensuremath{E\subTE}}
\def\ZTM{\ensuremath{Z\subTM}}
\def\VTM{\ensuremath{V\subTM}}

\def\etal{\textit{et.~al}}

\def\arreq{&{}={}&\ds }
\def\arrnone{&&\ds }
\def\arrap{&{}\approx{}&\ds }

\def\eqnarr#1#2{  
\renewcommand{\arraystretch}{#1}
  \setlength\arraycolsep{0ex}
  \begin{array}{rcl}
    #2
  \end{array}
}
\def\expct#1{\!\left\langle{#1}\right\rangle}
\def\Tparam{\cT}

\definecolor{darkblue}{rgb}{0.0,0.0,0.5}
\usepackage[colorlinks,citecolor=darkblue,linkcolor=darkblue]{hyperref}

\begin{document}

\author{Jonathan B. Mackrory}
\affiliation{Department of Physics and Oregon Center for Optical, Molecular and Quantum Science, 
1274 University of Oregon, Eugene, Oregon 97403-1274}
\author{Tanmoy Bhattacharya}
\affiliation{Santa Fe Institute, Santa Fe, NM 87501}
\affiliation{Los Alamos National Laboratory, Theoretical Division T-2, Los Alamos, NM 87545}
\author{Daniel A. Steck}
\affiliation{Department of Physics and Oregon Center for Optical, Molecular and Quantum Science, 
1274 University of Oregon, Eugene, Oregon 97403-1274}

\title{Worldline approach for numerical computation of electromagnetic Casimir energies.\\ 
  I. Scalar field coupled to magnetodielectric media}
\date{\today}

\begin{abstract}
  We present a worldline method for the calculation of Casimir energies for 
  scalar fields coupled to magnetodielectric media.  
  The scalar model we consider may be applied in arbitrary geometries, 
  and it corresponds exactly to one polarization of the electromagnetic 
  field in planar layered media.  
  Starting from the field theory for electromagnetism, we work with the two 
  decoupled polarizations in planar media and develop worldline path integrals, 
  which represent the two polarizations separately, for computing both 
  Casimir and Casimir--Polder potentials.  
  We then show analytically that the path integrals for the 
  transverse-electric (TE)  polarization coupled to a dielectric medium
  converge to the proper solutions in certain special cases, 
  including the Casimir--Polder potential of an atom near a planar interface, 
  and the Casimir energy due to two planar interfaces.  
  We also evaluate the path integrals numerically via Monte-Carlo path-averaging
  for these cases, studying the convergence and performance of the   resulting 
  computational techniques. 
  While these scalar methods are only exact in particular geometries, 
  they may serve as an approximation for Casimir energies for the vector 
  electromagnetic field in other geometries.
\end{abstract}

\maketitle

\section{Introduction}

The Casimir force is a striking manifestation of the quantum vacuum.  
Casimir forces arise due to fluctuations in quantum fields interacting with 
material bodies.  
These fluctuations lead to forces between ideal conductors~\cite{Casimir1948a}, 
atoms and conductors~\cite{CasimirPolder1948}, and dielectric slabs~\cite{Lifshitz1956}.
Beyond their fundamental interest as inherently quantum phenomena, 
Casimir forces are also of technical importance. 
For example, they must be taken into account in stiction in 
micro-electromechanical systems~\cite{Buks2001}, and in attempts to couple 
atoms to solid-state systems to realize architectures for 
a scalable quantum-information processor~\cite{Alton2010,Hung2013}.
It is then imperative to be able to calculate these forces in a 
wide range of geometries, while also taking into account material properties.  
The calculation of Casimir forces is challenging because the force depends 
sensitively on the material properties and the geometry of the bodies.
In addition, 
the calculations are usually expressed as differences between 
divergent quantities, which must be handled carefully.
While analytical calculations can be carried out for simple, highly symmetric geometries,
in general numerical approaches are required~\cite{Johnsonbook2011}.

The scattering approach is to date the only general method for calculating 
electromagnetic Casimir energies in arbitrary geometries of material bodies.
This method has been developed by a number of authors as an analytical 
tool~\cite{Lambrecht2006,Kenneth2006,Emig2007,MaiaNeto2008,Rahi2009}.
The scattering approach has also been extended to a general numerical method for 
computing Casimir energies~\cite{Rodriguez2007, Rodriguez2009,Reid2009, Reid2011}
by leveraging the computational similarity to calculations in classical 
electromagnetism~\cite{Johnsonbook2011}.
This ``boundary element method'' considers fluctuating surface currents on 
bodies interacting via the electromagnetic field.
Computationally, this method relies on inverting the matrix that 
encodes the scattering of photons by the surface currents~\cite{Reid2013}.
While indeed being powerful and general, it is important to complement this 
method with alternate methods, which would possess different systematic errors 
and alternative regimes of efficient operation.

The worldline method is a promising alternative method for calculating 
Casimir energies~\cite{Gies2003}. 
The world\-line method is a general method of computing effective actions 
for quadratic
quantum field theories in terms of worldline (i.e., single-particle) 
path integrals~\cite{Strassler1992, Schubert2001}.
Gies \etal\ showed how to apply the worldline method to computing Casimir energies
 for a scalar field coupled to a background potential, which models the material 
bodies~\cite{Gies2003,Gies2006,Gies2006a,Gies2006c,Gies2006d}.
More recently, this method has been extended to computing stress-energy
tensors for scalar fields, with applications to computing Casimir energies~
\cite{Schafer2012, Schafer2016}. 
Since the formalism is not specific to any particular geometry, it serves as a 
method for numerically computing Casimir energies in \textit{arbitrary} geometries.
Furthermore,  the worldline method offers an intuitive picture of Casimir 
energies as emerging from the spacetime trajectories (worldlines) of virtual particles.

In brief, in the worldline method, one generates an ensemble of closed, random walks, 
and along the walk one evaluates the potential, which encodes the locations 
and geometries of the interacting bodies. 
Thus, visualizing the intersection of the ``path cloud'' with the material 
bodies provides the intuitive picture of the Casimir energy.
For example, the worldline method has been applied in evaluating Casimir 
energies for nontrivial geometries, notably for a piston in a 
flasklike container~\cite{Schaden2009}, where the worldlines give an intuitive 
picture of how the piston and flask contributions conspire to produce 
a force whose sign depends on the shape of the flask. 
Worldline numerics have also been used to better understand geometries with 
sharp edges~\cite{Gies2006c}, and to understand the limits of the 
proximity-force approximation~\cite{Gies2006b}.  
Finally, the method has also been extended to include the effects of 
nonzero temperature~\cite{Klingmuller2008,Weber2009,Weber2010}.  

At present there are two main limitations of the worldline method.  
First, the method only treats a single scalar field, whereas two coupled 
polarizations are necessary to treat full electromagnetism.  Second, the 
material bodies are treated as a background potential, rather than a 
dielectric permittivity and a magnetic permeability.

The main success of the worldline method thus far has been in modeling bodies 
via arbitrarily strong potentials.
These result in Dirichlet boundary conditions on the scalar field, 
which mimic the boundary conditions on one electromagnetic polarization 
at a perfectly conducting boundary.
However, in extending the utility of the worldline method, it is important to 
treat the vector nature of the field and the coupling to magnetodielectric materials.

There has already been some progress in this direction thus far. 
  Bordag \etal\ have developed the path-integral  quantization of the 
electromagnetic field, including coupling to a nondispersive 
  dielectric~\cite{Bordag1998,Bordag1999}, with applications to the analytic 
computation of energies for dielectric spheres and heat-kernels coefficients.  
  Aehlig \etal, in a paper discussing computational optimizations to worldline 
calculations of Casimir energies, have also speculated on how polarization 
could be incorporated~\cite{Aehlig2011}.
  Fosco \etal\ have considered how to implement regularized Neumann boundary 
conditions in path integrals, and wrote down a worldline path integral 
for Neumann boundaries, using a nonlocal (in proper time) representation of the
 boundary potential~\cite{Fosco2010}.
Similar functional-determinant methods have also been used to compute
 the electrostatic contribution of the Casimir energy in terms of a 
scalar field~\cite{Pasquali2008}.  

In the present work we show how to incorporate explicit coupling to dielectric
 and magnetic materials within the worldline formalism. 
We demonstrate analytically how, in simple geometries, a new version of the
 scalar theory reproduces known  electromagnetic Casimir and Casimir--Polder
 energies for dielectric bodies.
We also study the numeric evaluation of the worldline path integrals. 
In this work we focus mainly on the path integral for the transverse-electric
(TE) polarization.
In the limit of large dielectric permittivity, this path integral also handles
 Dirichlet boundary conditions on the (scalar) TE field.
While the path integral can be evaluated in any geometry, it only corresponds 
exactly to one electromagnetic polarization in planar, layered media.
In other geometries, it can serve as a scalar approximation 
(albeit an uncontrolled one) for the full electromagnetic problem, 
including proper dielectric coupling.
The other, transverse magnetic (TM) polarization has additional technical 
complications for dielectric media, as it involves a singular potential 
at a dielectric interface.
In the limit of large dielectric permittivity, it imposes Neumann boundary 
conditions on the scalar field.  
We will discuss the analytic and numerical evaluation of this path integral 
in a future paper~\cite{TMpaper}.

This paper is organized as follows. 
In Sec.~\ref{sec:Partition_Function} we compare the action for the original 
scalar problem with the action for the electromagnetic field in media and 
introduce a scalar action to model the electromagnetic problem.
We then derive the corresponding partition function.
In Sec.~\ref{sec:worldline_path_integral} we develop worldline path 
integrals for the Casimir and Casimir--Polder energies.
In Sec.~\ref{sec:Analytical} we derive analytical results for the 
Casimir--Polder potential for an atom near a planar, dielectric interface, 
and the Casimir energy for two parallel, planar dielectric interfaces.
In Sec.~\ref{sec:Numerical} we discuss the numerical methods we use to 
evaluate the path integral, and finally in Sec.~\ref{sec:Thermal} we generalize
 the formalism to incorporate dispersion and nonzero temperature.  

\section{Field partition function}
\label{sec:Partition_Function}

To begin our development, we will briefly review the setup of the previous 
worldline formalism for evaluating the Casimir energy of a scalar field 
$\phi=\phi(\vect{r},t)$ coupled to a background potential $V(\vect{r})$.
This is described by the action~\cite{Gies2003}
\begin{equation}
  S = \frac{1}{2}\int_0^T dt \int d\vect{r}\left[
    \frac{1}{c^2}\left(\partial_t \phi\right)^2-|\nabla\phi|^2
    - V(\vect{r})\,\phi^2\right],
\end{equation}
with associated wave equation
\begin{equation}
  \nabla^2\phi-\frac{1}{c^2}\partial_t^2\phi+ V(\vect{r})\,\phi=0.
\end{equation}
In the original work~\cite{Gies2003}, the suggestion was to represent the 
potential in terms of a delta function as 
$V(\mathbf{r})=\lambda \,\delta\left[\sigma(\vect{r})\right]$, 
where $\sigma(\vect{r})=0$ defines the surfaces of the material bodies.
  An alternative representation arises by simply setting $V(\mathbf{r})=\lambda$
 in the interior of the bodies.  In the limit $\lambda\longrightarrow\infty$,
 the potential in either case imposes Dirichlet boundary conditions on the surfaces. 

In electromagnetism, by contrast, the source-free field action in the presence 
of linear, nondispersive media, in terms of the scalar and vector potentials 
$A_0$ and $\vect{A}$ is
\begin{align}
  S\subEM = \frac{1}{2\mu_0}\!\int\! dt \!\int\! d\vect{r} 
  \left[\frac{\epsr(\vect{r})}{c^2}(\nabla A_0+\partial_t\vect{A})^2
    - \frac{(\nabla\times\vect{A})^2}{\mur(\mathbf{r})}\right]\!,
  \label{eq:EM_action}
\end{align}
where $\epsr(\vect{r}):=\epsilon(\vect{r})/\epsilon_0$ and 
$\mur(\vect{r}):=\mu(\vect{r})/\mu_0$ are the relative permittivity 
and permeability, respectively.  
Note that this action may be equivalently written
\begin{equation}
  S\subEM= \frac{1}{2}\int dt \int d\vect{r}\, 
  \big(\vect{E}\cdot\vect{D}-\vect{B}\cdot\vect{H}\big),
\end{equation}
where the fields are given as usual by 
$\vect{E}=-(\nabla A_0+\partial_t\vect{A})$, $\vect{B}=\nabla\times\vect{A}$,
${\vect{D}(\vect{r},t)=\epsilon(\vect{r})\vect{E}(\vect{r},t)}$,  
and ${\vect{B}(\vect{r},t)=\mu(\vect{r})\vect{H}(\vect{r},t)}$.

\begin{figure}
  \includegraphics[width=0.75\columnwidth]{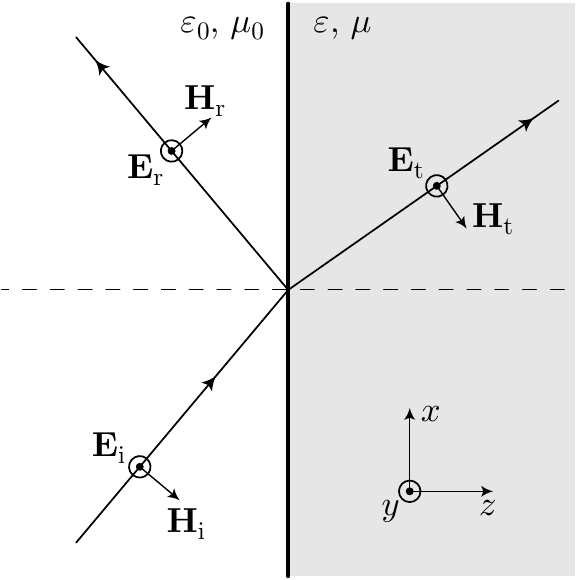}
  \caption{Schematic illustration of electric and magnetic fields 
   for a TE-polarized mode at a planar vacuum--dielectric interface,
   for incident ($\mathbf{E}_\mathrm{i}$, $\mathbf{H}_\mathrm{i}$),   
   reflected ($\mathbf{E}_\mathrm{r}$, $\mathbf{H}_\mathrm{r}$),   and
   and transmitted ($\mathbf{E}_\mathrm{t}$, $\mathbf{H}_\mathrm{t}$) fields. 
   For this polarization, the electric field is effectively scalar, while
   the magnetic-field  polarization varies between media.
   This scalar-like behavior holds in any planar layered medium that varies
   in only one direction, such that
   $\epsilon(\mathbf{r})\equiv\epsilon(z)$ and 
   $\mu(\mathbf{r})\equiv\mu(z)$.
   }
  \label{fig:TE_polarization}
\end{figure}

Then setting $\delta S\subEM/\delta A_0=0$ leads to 
$\nabla\cdot\epsr \nabla A_0=-\partial_{t}\nabla\cdot\epsr\mathbf{A}$.
This is a first-order equation in time, thus acting as a constraint
and implying a gauge freedom.
In Coulomb gauge, $\nabla\cdot\epsr\mathbf{A}=0$, 
and thus in the absence of sources we may take $A_0=0$.  
The remaining variation $\delta S\subEM/\delta \mathbf{A}=0$ 
leads to the wave equation
\begin{equation}
  \nabla\times\frac{1}{\mur}\nabla\times\vect{A} 
-\frac{\epsr}{c^2}\partial_t^2\vect{A} = 0.\label{eq:EM_field_equation}
\end{equation}
In the case of a planar layered medium such that the 
electromagnetic properties of the media only vary 
in one direction, $\epsr(\mathbf{r})\equiv\epsr(z)$ and $\mur(\mathrm{r})\equiv\mur(z)$,
 the action (\ref{eq:EM_action}) factors into two independent scalar actions, 
corresponding to the transverse-electric (TE) and transverse-magnetic (TM)
 polarizations.
This decomposition of the electromagnetic field into two decoupled 
scalar fields was also used by Schwinger \etal~\cite{Schwinger1978}
 for computing Casimir energies in planar geometries.
This is illustrated for the TE polarization at a planar interface in 
Fig.~\ref{fig:TE_polarization}, where for plane-wave modes, 
the electric-field component parallel to the medium behaves as a scalar,
 since its direction is fixed. 

In any  TE-polarized plane-wave mode, the electric field is parallel 
to the interface. We take the polarization to be along the $y$-direction, 
such that $\mathbf{E}=\hat{y}E$, and thus $\mathbf{A}=\hat{y}A$.
The Coulomb-gauge condition further implies $\partial_y{A}=0$, 
so that $(\nabla\times\vect{A})^2=|\nabla A_y|^2$.  
Then for any planar TE mode, the action (\ref{eq:EM_action}) can be expressed
in terms of the nonvanishing component of the 
vector potential $\phi(\vect{r},t):=A_y$ as
\begin{equation}
  S\subTE = \frac{1}{2\mu_0}\int dt \int d\vect{r} \left[
    \frac{\epsr(z)}{c^2}\left(\partial_t\phi\right)^2
    - \frac{1}{\mur(z)}|\nabla\phi|^2\right],
  \label{eq:TE_scalar_action}
\end{equation}
with corresponding wave equation 
\begin{equation}
  \nabla\cdot\frac{1}{\mur({z})}\nabla\phi 
  - \frac{\epsr({z})}{c^2}\partial_t^2\phi = 0.
\end{equation}
This action allows the TE-polarized modes to be
represented explicitly in terms of a scalar field $\phi(\vect{r},t)$.

To treat the action for the transverse-magnetic (TM) polarization in a 
planar layered medium, it is convenient to introduce
magnetic potentials $C_0$ and $\vect{C}$~\cite{Bordag1999}, 
in terms of which the fields are $\vect{H}=(\nabla C_0+\partial_t\vect{C})$
 and $\vect{D}=(\nabla\times\vect{C})$.  
In terms of these potentials, the electromagnetic action is 
\begin{align}
  S\subEM = \frac{1}{2\epsilon_0}\!\int\! dt \!\int\! d\vect{r} 
  \left[\frac{\mur(\vect{r})}{c^2}(\nabla C_0+\partial_t\vect{C})^2
    - \frac{(\nabla\times\vect{C})^2}{\epsr(\mathbf{r})}\right],
 \label{eq:dual_EM_action}
\end{align}
after changing the overall sign. 
Assuming the Coulomb-like gauge condition
$\nabla\cdot\mur\mathbf{C}=0$, the reduction to a scalar action for TM 
modes follows in the same way, with resulting action
\begin{align}
  S\subTM = \frac{1}{2\epsilon_0}\int dt \int d\vect{r} 
  \left[\frac{\mur(z)}{c^2}\left(\partial_t\psi\right)^2 
    - \frac{1}{\epsr(z)}|\nabla\psi|^2\right],
 \label{eq:TM_scalar_action}
\end{align}
and corresponding wave equation 
\begin{equation}
  \nabla\cdot\frac{1}{\epsr({z})}\nabla\psi 
  - \frac{\mur({z})}{c^2}\partial_t^2\psi = 0,
\end{equation}
where for any plane-wave mode $\psi(\mathbf{r},t)$ is the only 
nonvanishing component of $\mathbf{C}$.
The TM scalar action and wave equation also follow simply from the TE case 
by noting that electromagnetism is invariant under the duality symmetry
$\vect{E}\longleftrightarrow\vect{H}$, $\vect{B}\longleftrightarrow\vect{-D}$,
 and $\epsilon\longleftrightarrow\mu$.  

The partition function for either scalar field is then a path integral over
 the fields in terms of the Euclidean action 
(i.e., with the replacement $t\rightarrow -i\hbar\beta$).  
For example, the TE path integral becomes
\begin{align}
  \ZTE =& \int D\phi\,\exp\left[-\frac{\epsilon_0 c}{2\hbar}
    \int_0^{\beta\hbar c}d\tau\int d\vect{r}\,\right.\nonumber\\
  &\left.\times\bigg(\epsr(\vect{r})|\partial_{\tau}\phi(\vect{r},\tau)|^2 
    + \frac{1}{\mur(\vect{r})}|\nabla\phi(\vect{r},\tau)|^2\bigg)\right],
  \label{eq:field_partition_function}
\end{align}
where $\tau:=\beta\hbar c$.  
Changing to a scaled field variable 
\begin{equation}
  \tilde{\phi}(\vect{r},t):= \frac{\phi(\vect{r},t)}{\sqrt{\mur(\vect{r})}}
  \label{eq:fieldrescale}
\end{equation} 
removes explicit spatial dependence from the gradient term, 
but the change in integration measure to $D\tilde{\phi}$ introduces a 
functional determinant involving $\mur$.
This determinant ultimately drops out of the final calculation, 
provided we calculate only physically relevant differences between configurations
amount to translations and rearrangements of the materials.
Note also that subtleties involved in Faddeev--Popov-type gauge-fixing 
do not arise here, because the gauge condition used here is linear in the fields.

Then carrying out the Gaussian integration over the field variables, 
and performing the analogous procedure in the TM case, 
the partition functions become
\begin{equation}
  \begin{array}{rcl}
  \ZTE \!\!&=&\displaystyle\!\! 
  \det\left[ -\frac{1}{2}\epsr(\vect{r})\mur(\vect{r})\partial_\tau^2
    - \frac{1}{2}\nabla^2 - \VTE(\vect{r})\right]^{-1/2}\\
  \ZTM \!\!&=&\displaystyle\!\! 
  \det\left[-\frac{1}{2}\epsr(\vect{r})\mur(\vect{r})\partial_\tau^2
    - \frac{1}{2}\nabla^2 - \VTM(\vect{r})\right]^{-1/2}\!\!,
  \end{array}
  \label{eq:Z-det}
\end{equation}
where the potentials are defined as
\begin{equation}
  \renewcommand{\arraystretch}{2.1}
  \begin{array}{rcl}
  \VTE \!\!&:=&\displaystyle\!\!  
  \frac{1}{2}\left[(\nabla\log\sqrt{\mur})^2 -\nabla^2\log\sqrt{\mur}\right]\\
  \VTM \!\!&:=&\displaystyle\!\!  
  \frac{1}{2}\left[(\nabla\log\sqrt{\epsr})^2 -\nabla^2\log\sqrt{\epsr}\right]
  \end{array}
  \label{eq:TE_pot}
\end{equation}
and these arise from the commutation of $\mur(\vect{r})$ and $\epsr(\vect{r})$ 
through the derivative operators.
Note that $\VTE$ and $\VTM$ will still appear in the path integral
 even without the field-rescaling 
procedure in Eq.~(\ref{eq:fieldrescale}).  
However, the development of the worldline path integrals is much more involved. 
Further details on this alternate derivation of the 
potentials will be provided elsewhere~\cite{TMpaper}, 
where we will also discuss their analytical and numerical evaluation at material interfaces.

\section{Worldline Path Integral}
\label{sec:worldline_path_integral}

\subsection{Casimir energies and renormalization}

To extract zero-temperature Casimir energies from the partition function, 
we can compute the mean ground-state energy via $E=-\partial_\beta\log Z$.
Since the divergent \textit{absolute} energy is not a physical observable,
it is important to renormalize
 the energy. That is, one should subtract the energy of a reference configuration to 
obtain a finite interaction energy: $E=-\partial_\beta(\log Z-\log Z_0)$.
The reference configuration depends on the geometry of interest, 
but a typical choice is an arbitrarily large separation of the material objects.

The derivation of the worldline path integral then proceeds by evaluating the
partition-function determinants (\ref{eq:Z-det})
via the identity
\begin{equation}
  \log\det[A]= \tr\log[A],
\end{equation}
and then using the integral representation 
\begin{equation}
  \log[A]-\log[B] = \int_0^\infty\! \frac{d\cT}{\cT}\Big( e^{-B\cT}- e^{-A\cT}\Big)
\end{equation}
for the logarithm. 
Note that without the renormalizing subtraction here, the integral diverges
at the lower limit---this is the same ultraviolet divergence that comes from
 computing Casimir energies via mode-summation.
For simplicity of notation, we will normally not write out the subtraction 
of the reference configuration,
so the divergent expressions that follow must be interpreted as representing
the finite quantity that remains after this subtraction.
From Eq.~(\ref{eq:Z-det}), the (non-renormalized) mean energy is  
\begin{equation}
  \ETE = -\frac{\partial_\beta}{2}\!\!\int_0^\infty\!\! \frac{d\cT}{\cT}\, 
  \tr\, e^{\cT\left[(1/2)\epsr(\hat{\vect{x}})\mur(\hat{\vect{x}})\partial_\tau^2
      + (1/2)\nabla^2+\VTE(\hat{\vect{x}})\right]},
  \label{ETEtrace}
\end{equation}
where now the notation $\hat{\vect{x}}$ emphasizes the operator nature 
of the field variable.
We can evaluate the trace for this operator by introducing an auxiliary Hilbert
 space with 
 $\langle \vect{x},\tau| \op{\vect{x}}|\psi\rangle = \vect{x}\psi(\vect{x},\tau)$,
 $\langle \vect{x},\tau|\op{\vect{p}}|\psi\rangle = -i \nabla\psi(\vect{x},\tau)$,
 $\langle \vect{x},\tau| \op{\tau}|\psi\rangle = \tau\psi(\vect{x},\tau)$,
 $\langle \vect{x},\tau|\op{p}_\tau|\psi\rangle = -i\partial_\tau\psi(\vect{x},\tau),$
 and $\langle \vect{x}|\vect{p} \rangle =e^{i\vect{x}\cdot\vect{p}}$.
Then expressing the trace as a spacetime integration, the result is
\begin{align}
  \ETE &= -\frac{\partial_\beta}{2}\int_0^\infty\!\! \frac{d\cT}{\cT}\,
  \int_0^{\beta\hbar c}\!\!\!d\tau_0 \int d\vect{x}_0 \nonumber\\
  &\times\langle \vect{x}_0,\tau_0|e^{-\left(\epsr(\op{\vect{x}})\mur(\op{\vect{x}})\op{p}_\tau^2/2
      +\op{\vect{p}}^2/2+\VTE(\op{\vect{x}})\right)\cT}|\vect{x}_0,\tau_0\rangle.
  \label{eq:path_int_operator}
\end{align}
Since the matrix element is independent of $\op{\tau}$, 
it is only necessary to develop the path integral in the spatial dimensions.
Splitting the exponential operator into $N$ pieces, 
and inserting spatial position and momentum identities 
$1=\int d\mathbf{x}_j\,|\mathbf{x}_j\rangle\langle\mathbf{x}_j|=
(2\pi)^{-(D-1)}\int d\mathbf{p}_j\,|\mathbf{p}_j\rangle\langle\mathbf{p}_j|$
between each piece,
the matrix element becomes   
\begin{align}
  &\hspace{-6mm}\langle \vect{x}_0,\tau_0|e^{-\left[\epsr(\op{\vect{x}})\mur(\op{\vect{x}})\op{p}_\tau^2/2
      +\op{\vect{p}}^2/2+\VTE(\op{\vect{x}})\right]\cT}|\vect{x}_0,\tau_0\rangle \nonumber \\
  \hspace{2mm}= &\int \prod_{j=1}^{N}\left(\frac{d\vect{x}_jd\vect{p}_j}{(2\pi)^{D-1}}\right)
  \bigg\{ \delta^{D-1}(\vect{x}_N-\vect{x}_0)  \prod_{k=0}^{N}\big[\langle \vect{x}_{k} |\vect{p}_{k}\rangle \nonumber \\
  &\times\langle \vect{p}_{k}|e^{-\left(\epsr(\op{\vect{x}})\mur(\op{\vect{x}})\op{p}_\tau^2/2
      +\op{\vect{p}}^2/2+\VTE(\op{\vect{x}})\right)\Delta\cT}|\vect{x}_{k-1}\rangle\big]\bigg\},
  \label{eq:path_int_op}
\end{align}
where $D$ is the spacetime dimension, $\Delta\cT=\cT/N$, and the path-closure
 condition $\vect{x}_N=\vect{x}_0$ from the trace is now enforced by the 
delta function.  
Replacing operators by eigenvalues,
using $\langle \mathbf{x}_j|\mathbf{p}_j\rangle=e^{i\mathbf{p}_j\cdot\mathbf{x}_j}$,
 and evaluating the momentum integrals leaves
\begin{align}
  &\hspace{-6mm}\langle \vect{x}_0,\tau_0|e^{-\left[\epsr(\op{\vect{x}})\mur(\op{\vect{x}})\op{p}_\tau^2/2
      +\op{\vect{p}}^2/2+\VTE(\op{\vect{x}})\right]\cT}|\vect{x}_0,\tau_0\rangle \nonumber \\
  \hspace{2mm}= &\int \prod_{j=1}^{N}\frac{d\vect{x}_j}{(2\pi\Delta \cT)^{(D-1)/2}}
  \bigg\{\delta^{D-1}(\vect{x}_N-\vect{x}_0)\nonumber\\
  &\times\prod_{k=1}^{N}\left[e^{-(\vect{x}_k-\vect{x}_{k-1})^2/(2\Delta \cT)
      -\Delta\cT \VTE(\vect{x}_{k-1})}\right.\nonumber\\
  &\left.\times \,e^{-\epsr(\vect{x}_{k-1})\,\mur(\vect{x}_{k-1})\,\op{p}_\tau^2\Delta \cT/2}
  \right]\bigg\}.
  \label{eq:path_int_eig}
\end{align}
Putting this back into Eq.~(\ref{eq:path_int_operator}) and carrying out
 the remaining integrals over $\tau_0$ and  $p_\tau$ gives
\begin{align}
  \ETE = &-\frac{\hbar c}{2}\int_0^\infty\!\! \frac{d\cT}{\cT}
  \int \prod_{j=1}^{N}\frac{d\vect{x}_j}{(2\pi\Delta \cT)^{(D-1)/2}}\nonumber
  \\&\times 
  \frac{\delta^{D-1}(\vect{x}_N-\vect{x}_0)}{[2\pi \cT N^{-1}\sum_{\ell=0}^{N-1}\epsr(\vect{x}_\ell)\,\mur(\vect{x}_\ell)]^{1/2}}
  \nonumber\\
  &\times \prod_{k=1}^{N}\left[e^{-(\vect{x}_k-\vect{x}_{k-1})^2/(2\Delta \cT)-\Delta\cT \VTE(\vect{x}_{k-1})}\right].
  \label{discretepathintegralwithmeasure}
\end{align}
This expression already exhibits the form of the numerical method: 
The Gaussian densities, in conjunction with the delta function, 
define a probability measure for paths (random walks) that begin
 and end at $\vect{x}_0$. 
The contribution of the material body enters in the evaluation
 of $[\epsr(\vect{r})\,\mur(\vect{r})]$ and $\VTE(\vect{r})$ 
[or $\VTM(\vect{r})$ in the TM case] along the path.

The presence of the delta function $\delta(\vect{x}_N-\vect{x}_0)$ 
leads to an additional overall normalization constant $(2\pi\cT)^{-(D-1)/2}$,
if the sum encompasses only paths that close, such that $\vect{x}_N=\vect{x}_0$.
This follows from the expression~\cite{Hormander1983} 
\begin{equation}
  \int d\mathbf{q}\, \delta[h(\vect{q})]\,f(\vect{q})
  = \int_{h^{-1}(0)} \!\!\!\!\!dS\,\frac{1}{|\nabla h(\vect{q})|}f(\vect{q}),
  \label{hormander1}
\end{equation}
where $S$ is the surface satisfying $h(\vect{q})=0$, and 
\begin{equation}
|\nabla h(\vect{q})|=\sqrt{\sum_k \left(\frac{\partial h}{\partial q_k}\right)^2}
  \label{hormander2}
\end{equation}
is the Euclidean norm of the gradient vector.
In the case at hand, the delta function restricts a sum of $N$ Gaussian integrals to have
a total of zero.
Defining the increments $\Delta\vect{x}_n := \vect{x}_{n+1}-\vect{x}_n$, 
the path-closure constraint is $\smash{\delta(\sum_{k=0}^{N-1} \Delta\vect{x}_k)}$.
Accounting for the remaining normalization constant of the Gaussian 
$\vect{x}_N$ integral (as part of $dS$), the extra contribution for 
considering only closed paths ($\vect{x}_N=\vect{x}_0$)  in the path integral
 is $(2\pi\Delta \cT N)^{-(D-1)/2} = (2\pi\cT)^{-(D-1)/2}$.  

Taking the remaining $N-1$ Gaussian factors to be the probability measure 
for the paths, in a Monte-Carlo interpretation of the path integrals,
 the sample paths are Brownian bridges~\cite{Karatzas1991} 
in the limit of large $N$. 
To be more precise, a Wiener path $W(t)$ is a continuous time random walk,  
where each increment has ensemble average $\dlangle dW(t)\drangle =0$ 
(where the double brackets denote an ensemble average) and
variance $\dlangle dW^2(t)\drangle = dt$, 
with $\dlangle dW(t)\,dW(t')\drangle=0$ for $t\neq t'$.
A Brownian bridge $B_\cT(t)$ is a Wiener path subject to the pinning condition 
$B_\cT(\cT)=0$.  
(The bridges may also be defined such that they are pinned to other endpoints.)
Brownian bridges thus form a subset of zero measure of all possible Wiener paths.

The (unrenormalized) Casimir energy can then be rewritten in continuous-time notation as 
\begin{equation}
  E\subTE = -\frac{\hbar c}{2(2\pi)^{D/2}}\int_0^\infty\!\!\! \frac{d\cT}{\cT^{1+D/2}} 
\int\! d\vect{x}_0\,\Biggdlangle \frac{e^{-\cT\langle \VTE\rangle}}{\langle\epsr\mur\rangle^{1/2}}\Biggdrangle_{\!\!\vect{x}(t)}
  \label{eq:Casimir worldline}
\end{equation}
for the TE polarization, and
\begin{equation}
  E\subTM = -\frac{\hbar c}{2(2\pi)^{D/2}}\int_0^\infty\!\!\! \frac{d\cT}{\cT^{1+D/2}} 
\int\! d\vect{x}_0\,\Biggdlangle \frac{e^{-\cT\langle \VTM\rangle}}{\langle\epsr\mur\rangle^{1/2}}\Biggdrangle_{\!\!\vect{x}(t)}
  \label{eq:Casimir worldlineTM}
\end{equation}
for the TM polarization,
where only the coordinate $\mathbf{x}_0$ is elided
in the notation $\smash{\dlangle \cdots\drangle_{\vect{x}(t)}}$ that denotes an ensemble average over 
vector Brownian bridges 
$\vect{x}(t)$ starting and returning to $\vect{x}_0$,
 and 
\begin{equation}
  \langle f \rangle := \frac{1}{\cT}\int_0^\cT \!\!dt\, 
  f[\vect{x}(t)]=\frac{1}{N}\sum_{k=0}^{N-1}f(\vect{x}_k)\label{eq:path_average}
\end{equation}
is a shorthand for the average value of $f(\mathbf{r})$, evaluated along the path.
In this paper, we will stick to 
the evaluation of the TE path integral (\ref{eq:Casimir worldline})
in the presence of dielectric-only materials, which simplifies to
\begin{equation}
  E\subTE = -\frac{\hbar c}{2(2\pi)^{D/2}}\int_0^\infty\!\!\! \frac{d\cT}{\cT^{1+D/2}} 
\int\! d\vect{x}_0\,\Bigdlangle \langle\epsr\rangle^{-1/2}\Bigdrangle_{\!\!\vect{x}(t)}.
  \label{eq:Casimir worldline TEdiel}
\end{equation}
Note that this has the same form as the TM path integral (\ref{eq:Casimir worldlineTM})
in the presence of magnetic-only materials.

\subsection{Casimir--Polder energies}
\label{sec:Casimir--Polder}

In principle, Eqs.~(\ref{eq:Casimir worldline}) and (\ref{eq:Casimir worldlineTM})
can yield Casimir--Polder energies for an atom interacting with a macroscopic body
by treating the atom as a small chunk of magnetodielectric material.
However, this is numerically inefficient: the vast majority of paths will not
intersect the atom, and will thus not contribute to the renormalized potential.
(Note that only paths that intersect both bodies will contribute to the 
properly renormalized, two-body Casimir energy.)
It is therefore useful to develop specialized path integrals for evaluating
Casimir--Polder potentials.

To introduce the atom, we may regard it as producing
localized perturbations $\delta\epsr(\mathbf{r})$ and $\delta\mur(\mathbf{r})$
to the background relative permittivity and permeability, respectively.
In the dipole approximation, 
these perturbations are given explicitly in terms of delta functions as
\begin{equation}
  \renewcommand{\arraystretch}{2.3}
  \begin{array}{rcl}
  \delta\epsr(\vect{r})\!\!&=&\!\!\ds\frac{\alpha_0}{\epsilon_0}\delta^{D-1}(\vect{r}-\rA)\\
  \delta\mur(\vect{r}) \!\!&=&\!\!\ds\beta_0\mu_0\,\delta^{D-1}(\vect{r}-\rA),
  \end{array}
  \label{epsmuperturbations}
\end{equation}
where $\alpha_0$ and $\beta_0$ are respectively the static polarizability
 and magnetizability of the atom, and $\rA$ is the location of the atom.
[Note that the polarizability and magnetizability are defined such that 
the electric and magnetic induced dipole moments are 
$\mathbf{d}=\alpha_0\mathbf{E}$ and $\mathbf{m}=\beta_0\mu_0\mathbf{H}$, 
while the polarization and magnetization are conventionally given in terms 
of the perturbations by $\mathbf{P}=\delta\epsilon\,\mathbf{E}$ and
$\mathbf{M}=(\delta\mu/\mu_0)\,\mathbf{H}$.]
The relevant energy is then the difference between the Casimir energies with 
and without the perturbations.  
To first order in the perturbations,
\begin{equation}
  \begin{array}{rcl}
  \delta E[\epsr,\mur]
  \!\!&=&\!\!\ds E[\epsr+\delta\epsr,\mur+\delta\mur] -E[\epsr,\mur] \\
  \!\!&=&\!\!\ds \int d\mathbf{r}\left[
    \frac{\delta E}{\delta\epsr}\,\delta\epsr(\mathbf{r}) 
    + \frac{\delta E}{\delta\mur}\,\delta\mur(\mathbf{r})\right].
  \end{array}
  \label{deltaEmueps}
\end{equation}
Then putting in the perturbations (\ref{epsmuperturbations}), 
we identify this energy as the Casimir--Polder potential,
\begin{equation}
  V\subCP(\rA)
  = \frac{\alpha_0}{\epsilon_0}\left(\frac{\delta E}{\delta\epsr(\rA)}\right) 
  + \beta_0\mu_0\left(\frac{\delta E}{\delta\mur(\rA)}\right),
  \label{VCP}
\end{equation}
written here in terms of functional derivatives of the Casimir energy evaluated at the atomic position.
Note that in treating the atom as arbitrarily well-localized, 
the perturbations (\ref{epsmuperturbations})
are technically divergent. However, the \textit{effect} of the perturbations
is still small, so the expansion here is really in terms of $\alpha_0$ and $\beta_0$
(i.e., this is a shorthand for taking a small but finite radius of the atom to zero
at the end of the calculation).
Note also that the expression (\ref{VCP}) is invariant under the duality
transformation
$\epsr\longleftrightarrow\mur$, $\alpha_0/\epsilon_0\longleftrightarrow\beta_0\mu_0$
\cite{Buhmann2009,Safari2008}.
 
Then to carry out the explicit expansion in Eq.~(\ref{deltaEmueps}) to
compute the functional derivatives in the Casimir--Polder potential~(\ref{VCP}),
we need the path-average expansion
\begin{equation}
  \begin{array}{l}\ds
    \big\langle (\epsr+\delta\epsr)(\mur+\delta\mur)\big\rangle^{-1/2}
    \\\ds
    \hspace{15mm}= \frac{1}{\langle \epsr\mur\rangle^{1/2}}
    -\frac{\langle\delta\epsr\,\mu_r\rangle}{2\langle\epsr\mur\rangle^{3/2}}
    -\frac{\langle\epsr\,\delta\mur\rangle}{2\langle\epsr\mur\rangle^{3/2}},
  \end{array}
\end{equation}
as well as the expansions of the potentials (\ref{eq:TE_pot}), which are necessary
to include the contributions of the potential factors of the form $e^{-\cT\langle \VTE\rangle}$:
\begin{equation}
  \begin{array}{l}
  \VTE\big[\mur+\delta\mur\big]
  \\\hspace{5mm}=\ds\frac{1}{8}\big[\nabla\log(\mur+\delta\mur)\big]^2 
  - \frac{1}{4}\nabla^2\log(\mur+\delta\mur)\\
  \\\hspace{5mm}=\ds\VTE[\mur]+\frac{1}{4}\left[
    \left(\nabla\frac{\delta\mur}{\mur}\right) \cdot\nabla\log\mur
    - \nabla^2\frac{\delta\mur}{\mur}\right].
  \end{array}
\end{equation}
These expressions serve to expand Eqs.~(\ref{eq:Casimir worldline}) and (\ref{eq:Casimir worldlineTM}),
integrating by parts where necessary.
Integrating over the starting (and ending) point $\mathbf{x}_0$ of all rigid
 translations of a particular path and computing the path average gives
\begin{equation}
    \int d\mathbf{x}_0\,
    \big\langle f(\mathbf{x})\,\delta^{D-1}(\mathbf{x}-\rA)\big\rangle
    =
    f(\rA),
  \label{derivpathintvar3}
\end{equation}
where,
after removing the delta function,
the resulting ``paths'' 
are averages over all 
rigid translations of the same path such that the path intersects $\rA$.
The ensemble average of paths is equivalently sampled by simply averaging over paths
beginning and ending at $\rA$.
Assembling these pieces gives the following expression for the TE-polarization Casimir--Polder potential:
\begin{equation}
  \renewcommand{\arraystretch}{1.7}
 \begin{array}{l}\ds
   V\subCP\supTE(\rA) \\\ds
   \hspace{3mm}= \frac{\hbar c}{4(2\pi)^{D/2}}\int_0^\infty \!\!\frac{d\cT}{\cT^{1+D/2}} \\\ds
   \hspace{8mm}{}\times \Biggdlangle \left(
     \frac{\alpha_0\mur(\rA)}{\epsilon_0}+\beta_0\mu_0 \epsr(\rA)\right)
   \frac{e^{ -\cT\langle \VTE\rangle}}{\langle\epsr\mur\rangle^{3/2}}\\\ds
   \hspace{16mm}-\frac{\beta_0\mu_0\cT}{2\mur(\rA)}
   \Big[\nabla^2(\log\mur)+\nabla(\log\mur)\cdot\nabla +\nabla^2\Big] \\\ds
   \hspace{33mm} \times\frac{e^{ -\cT\langle \VTE\rangle}}{\langle\epsr\mur\rangle^{1/2}}
   \Biggdrangle_{\vect{x}(t),\,\mathbf{x}(0)=\mathbf{x}(\cT)=\rA}.
 \end{array}
  \label{VCPTE}
\end{equation}
For TM polarization, the corresponding expression is
\begin{equation}
  \renewcommand{\arraystretch}{1.7}
 \begin{array}{l}\ds
  V\subCP\supTM(\rA) \\\ds
  \hspace{3mm}= \frac{\hbar c}{4(2\pi)^{D/2}}\int_0^\infty \!\!\frac{d\cT}{\cT^{1+D/2}} \\\ds
  \hspace{8mm}{}\times \Biggdlangle \left(\beta_0\mu_0\epsr(\rA)+\frac{\alpha_0\mur(\rA)}{\epsilon_0} \right)
  \frac{e^{ -\cT\langle \VTM\rangle}}{\langle\epsr\mur\rangle^{3/2}}\\\ds
\hspace{16mm}-\frac{\alpha_0\cT}{2\epsilon_0\epsr(\rA)}
\Big[\nabla^2(\log\epsr)+\nabla(\log\epsr)\cdot\nabla +\nabla^2\Big] \\\ds
\hspace{33mm} \times\frac{e^{ -\cT\langle \VTM\rangle}}{\langle\epsr\mur\rangle^{1/2}}
\Biggdrangle_{\vect{x}(t),\,\mathbf{x}(0)=\mathbf{x}(\cT)=\rA}.
  \end{array}
  \label{VCPTM}
\end{equation}
In this paper we will evaluate the TE path integral for an electric-dipole atom coupled
to a dielectric-only material, in which case the path integral (\ref{VCPTE}) 
simplifies considerably to
\begin{equation}
  V\subCP\supTE(\rA) 
  = \frac{\hbar c \alpha_0}{4(2\pi)^{D/2}\epsilon_0}\int_0^\infty \!\!\frac{d\cT}{\cT^{1+D/2}} 
  \,\Bigdlangle \langle\epsr\rangle^{-3/2}
\Bigdrangle_{\vect{x}(t)},
  \label{VCPTEdielectric}
\end{equation}
which has the same form as the TM path integral for a magnetic-dipole atom 
coupled to a magnetic-only material.
The TM path integral also simplifies somewhat in this case, but still involves the
potential $V\subTM$ in Eqs.~(\ref{eq:TE_pot}), the evaluation of which we will consider
in the future~\cite{TMpaper}.

\section{Analytic Worldline Summation}\label{sec:Analytical}
  
To further investigate the worldline path integrals, we will consider
their analytic evaluation and show that the dielectric-body
path integrals (\ref{eq:Casimir worldline TEdiel}) and 
(\ref{VCPTEdielectric}) converge to known solutions in
planar geometries.
In certain limits, this evaluation is quite straightforward.
For example, for a polarizable atom interacting with a perfectly conducting planar surface
(corresponding to $\epsr\longrightarrow\infty$), the conductor imposes 
Dirichlet boundary conditions on the scalar field,
as in previous work on Casimir worldlines with background potentials~\cite{Gies2003}.
In the renormalized form of the path integral (\ref{VCPTEdielectric}),
the integrand $\langle \epsr\rangle^{-3/2}-1$ is averaged over the ensemble.
The integrand vanishes for paths that do not touch the surface,
but takes the value $-1$ for those that do. Thus, the 
ensemble average yields the probability for a Brownian bridge to
cross the surface, with an overall minus sign.
For a Brownian bridge $B_\cT(t)$, the probability to cross a boundary
at distance $d$ is well-known, and takes the value 
\begin{equation}
  P_\mathrm{cross}=\exp(-2d^2/\cT).
  \label{crossprob}
\end{equation}
Putting this value into the path integral and carrying out the remaining integral
gives
\begin{equation}
  V\subCP\supTE(\rA) 
  = -\frac{\hbar c \alpha_0}{64\pi^{2}\epsilon_0d^4}
  \label{VCPTEdielectric-perfect}
\end{equation}
for $D=4$, which is the correct contribution of the TE
polarization to the Casimir--Polder potential in this limit.

In the case of a more general planar dielectric surface,
the relevant statistic to describe the path
average $\langle \epsr\rangle$ is the sojourn-time functional
(see Appendix~\ref{appendix:sojourn})
\begin{equation}
  T_\mathrm{s}[B_\cT(t);d]:=\int_0^1\!dt\,\Theta[B_\cT(t)-d],
  \label{sojourndefB}
\end{equation}
which is the time a Brownian bridge $B_\cT(t)$ spends past
a boundary at distance $d$ (here $\Theta(x)$ is the Heaviside
function).  The probability distribution for $T_\mathrm{s}$
is known exactly for a Brownian bridge~\cite{Takacs1999,Hooghiemstra2002},
and the probability density may be written
\begin{equation}
  \renewcommand{\arraystretch}{1.8}
  \begin{array}{rcl}\ds
   f_{T_\mathrm{s}}(x)
    \!\!&=&\!\!\ds
   \left[1-e^{-2d^2/\cT}\right]\delta(x-0^+)
   \\&&\ds{}
   +
   \sqrt{\frac{8d^2(\cT-x)}{\pi x\cT^3}}
   \,e^{-2d^2/(\cT-x)}
   \\&&\ds{}
   +\left(1-\frac{4d^2}{\cT}\right)\frac{e^{-2d^2/\cT}}{\cT}
   \,\mathrm{erfc}\left(\sqrt{\frac{2d^2x}{\cT(\cT-x)}}\right)
   .
  \end{array}
  \label{sojournbridgefinal3}
\end{equation}
This may  be used to compute the ensemble average
of $\langle \epsr\rangle^{-3/2}-1=(1+\chi T_\mathrm{s})^{-3/2}-1$,
where $\chi=\epsr-1$ is the dielectric susceptibility,
in the renormalized form of the path integral (\ref{VCPTEdielectric})
which then yields the TE Casimir--Polder potential for arbitrary $\chi$.
However, we will defer this solution in favor of deriving it via a slightly
different method.

\subsection{Iterated Laplace transform}

The derivation of the sojourn-time density (\ref{sojournbridgefinal3})
involves the solution to a diffusion equation to obtain
an iterated Laplace transform of the density. Inverting the Laplace
transforms then gives the density directly \cite{Hooghiemstra2002}.
A modification of this procedure provides the same solution
for the Casimir--Polder potential for a polarizable atom
and a planar dielectric half-space without directly
knowing the sojourn-time density.  This method then extends
to other situations where the density for the
relevant path statistic has a cumbersome form
(such as the Casimir energy or Casimir--Polder potential
for two parallel, planar dielectric interfaces)
or is not readily available in closed form 
(such as the Casimir energies for the TM polarization for
planar dielectric interfaces).

The method employs the Feynman-Kac formula~\cite{Durrett1996,Karatzas1991}, 
which states that a solution $f(x,t)$ to the diffusion equation
\begin{equation}
  \partial_t f = \frac{1}{2}\partial_x^2 f  - [V(x)+\lambda]f +g(x),
\end{equation}
with particle killing rate $V(x)$ and source function $g(x)$,
can also be written as an ensemble average over diffusive trajectories,
\begin{align}
    f(x,t) = &\,\biggdlangle  f_0[x+W(t)]\, e^{-\lambda t - \int_0^t \!dt'\,V[x+W(t')]} \nonumber\\
&+ \int_0^t \!dt'\,g[x+W(t')] \,e^{-\lambda t'-\int_0^{t'} \!dt'' \,V[x+W(t'')]} \biggdrangle ,
\end{align}
where the initial condition is $f_0(x)=f(x,t=0)$, and 
$\dlangle \cdots\drangle$ denotes the ensemble average over 
Wiener processes $W(t)$ (with initial condition $W(0)=0$,
increment means $\dlangle dW(t)\drangle=0$ and increment variance
$\dlangle dW(t)^2\drangle = dt$).
The formula in one dimension is sufficient for Casimir calculations
with planar geometries, but the method here readily generalizes to multiple dimensions.

In steady state, the solution
is independent of the initial condition, and choosing the source
function to be $g(x,t)=\delta(x)$, the path average becomes
\begin{equation}
    f(x) = \int_0^\infty\!\! dt'\,\biggdlangle \delta[x+W(t')] \,
    e^{-\lambda t'-\int_0^{t'} \!dt''\, V[x(t'')]} \biggdrangle,
    \label{eq:path_int_solution}
\end{equation}
where $f(x)$ satisfies the eigenvalue equation
\begin{equation}
  \lambda f = \frac{1}{2}\partial_x^2 f  - V(x) \,f +\delta(x).
\label{eq:path_int_solution_ODE}
\end{equation}
The delta function pins the paths at the end point such that $W(t)=-x$.
In the worldline path integrals, solutions with
closed paths are
 then obtained by setting $x=0$ in the solution $f(x)$.
The further restriction to Brownian-bridge sample paths 
requires a further normalization factor of $\smash{(2\pi \cT)^{-1/2}}$, which follows
from Eqs.~(\ref{hormander1}) and (\ref{hormander2}).  

The worldline path integrals (\ref{eq:Casimir worldline TEdiel}) and 
(\ref{VCPTEdielectric}) for the Casimir and Casimir--Polder energies both have
the general form
\begin{equation}
  U({x}_0) = \int_0^\infty \!\!\frac{d\cT}{\cT^{1+D/2}}\,
  \biggdlangle \frac{1}{\langle\epsr\rangle^\alpha}\biggdrangle_{\!{x}(t)},
  \label{Ucasimir}
\end{equation}
where $\alpha=1/2$ for Casimir energies and $\alpha=3/2$ for Casimir--Polder energies,
and the paths $x(t)$ satisfy $x(0)=x(\cT)=x_0$.
Then the identity
\begin{equation}
  \frac{1}{[h({x})]^\alpha} = \frac{1}{\Gamma(\alpha)}\int_0^\infty \!\!ds\, s^{\alpha-1} e^{-s\,h({x})},
  \label{eq:inverse-moment}
\end{equation}
where $\Gamma[\alpha]$ is the gamma function,
allows us to exponentiate the material dependence in Eq.~(\ref{Ucasimir}):
\begin{align}
    U(x_0) =&\, \frac{1}{\Gamma(\alpha)}\int_0^\infty\!\! \frac{d\cT}{\cT^{1+(D-1)/2-\alpha}}
    \int_0^\infty \!\! ds\,s^{\alpha-1} \nonumber\\
    &\hspace{15mm}{}\times\biggdlangle \frac{1}{\sqrt{\cT}}\,e^{-s\int_0^\cT dt\,\epsr[x(t)]}\biggdrangle_{\!{x}(t)}.
\end{align}
The $\cT$ integral here has the form of a Mellin transform which can be
related to a Laplace transform via an integration of the form~\cite{Lew1975}
\begin{align}
  &\int_0^\infty \!\!d\cT\, \cT^{z-1}\,f(\cT) \nonumber \\
  &{}\hspace{5mm}= \frac{1}{\Gamma(1-z)}\int_0^\infty \!\!d\lambda\,\lambda^{-z}\int_0^\infty \!\!d\cT\, e^{-\lambda \cT}f(\cT)   .
  \label{eq:Laplace-Mellin}
\end{align}
Thus, the factor $\cT^{-1-(D-1)/2+\alpha}$ may be recast as an additional integral:
\begin{align}
    U(x_0) =& \int_0^\infty\!\! d\lambda\, \frac{\lambda^{(D-1)/2-\alpha}}{\Gamma[(D-1)/2-\alpha]}\int_0^\infty \!\!ds\,\frac{s^{\alpha-1}}{\Gamma(\alpha)}\nonumber\\
    &\hspace{3mm}{}\times\int_0^\infty \!\!d\cT\, \biggdlangle \frac{1}{\sqrt{\cT}}\,e^{-\lambda \cT-s\int_0^{\cT} dt\,\epsr[x(t)]}\biggdrangle_{\!{x}(t)}.
   \label{eq:Casimir-Feynman-Kac}
\end{align}
When written in terms of the susceptibility $\chi$, where $\epsr=1+\chi$, this expression 
takes the form of an iterated Laplace transform, in the variables $\lambda$ and $s$.
The solution $f(0)$ from Eq.~(\ref{eq:path_int_solution_ODE}) gives
the integral over the ensemble average here, and the remaining two integrals may then be carried out
to give the relevant Casimir energy.

\subsection{Casimir--Polder energy}\label{section:CPanalytic}

For an atom in vacuum, the explicitly renormalized form of the Casimir--Polder
potential (\ref{VCPTEdielectric}) is
\begin{equation}
  V\subCP\supTE(\rA) 
  = \frac{\hbar c \alpha_0}{4(2\pi)^{D/2}\epsilon_0}\int_0^\infty \!\!\frac{d\cT}{\cT^{1+D/2}} 
  \,\Bigdlangle \langle\epsr\rangle^{-3/2}-1
\Bigdrangle_{\vect{x}(t)}\!,
  \label{VCPTEdielectricrenorm}
\end{equation}
where the extra term corresponds to a subtraction of the potential in the limit where 
material bodies are moved far away from the atom.
This subtraction removes the $\cT=0$ divergence.
For an atom at distance $d$ from a planar dielectric half-space, 
we take $\epsr(z) = 1+\chi\Theta(z-d)$, where $d>0$.
The corresponding solution to Eq.~(\ref{eq:path_int_solution_ODE}) is
(see Appendix~\ref{Appendix:TE 1body})
\begin{align}
    &\int_0^\infty \!\!\frac{d\cT}{\sqrt{\cT}}\, \biggdlangle e^{-\lambda \cT-s \int_0^{\cT} dt\,(1+\chi \Theta[d-x(t)])}\biggdrangle_{x(t)} \nonumber\\
    &=\!\sqrt{\frac{\pi}{\lambda+s}}\left[1+ \frac{\sqrt{\lambda+s} - \sqrt{\lambda+s(1+\chi)}}{\sqrt{\lambda+s}+\sqrt{\lambda+s(1+\chi)}} \,e^{-2\sqrt{2(\lambda+s)}\,d}\right]\!.
    \label{eq:FK TE1}
\end{align}
This expression is related to the sojourn time of a Brownian bridge, and can be used to derive
the density (\ref{sojournbridgefinal3}).

The Casimir--Polder potential follows by substituting Eq.~(\ref{eq:FK TE1}) into
Eq.~(\ref{eq:Casimir-Feynman-Kac}), and then using the result with $\alpha=3/2$
to evaluate Eq.~(\ref{VCPTEdielectricrenorm}), with the result
in $D=4$ dimensions
\begin{align}
    V\subCP(d)=&-\!\frac{\hbar c\alpha_0}{8\pi^2\epsilon_0}
    \int_0^\infty \!\!ds\, \sqrt{s}\int_0^\infty\!\! d\lambda \nonumber\\
    &\hspace{2mm}{}\times\frac{e^{-2\sqrt{2(\lambda+s)}\,d}}{\sqrt{\lambda+s}} 
    \left(\frac{\sqrt{\lambda+s(1+\chi)}-\sqrt{\lambda+s}}{\sqrt{\lambda+s(1+\chi)}+\sqrt{\lambda+s}}\right),
  \label{VCPanalyticint}
\end{align}
where the $d$-independent part of Eq.~(\ref{eq:FK TE1}) 
vanishes under renormalization.
The integrals can be evaluated exactly to obtain
\begin{equation}
  V\subCP\supTE(d)  = -\frac{3\hbar c\alpha_0}{32\pi^2\epsilon_0 d^4}\,\eta\subTE(\chi),
  \label{VCPanalytic}
\end{equation}
where $\eta\subTE(\chi)$ gives the potential relative to the
Casimir--Polder energy between a spherical atom and a perfectly conducting plane:
\begin{equation}
  \eta\subTE(\chi)=
  \frac{1}{6} + \frac{1}{\chi} -\frac{\sqrt{1+\chi}}{2\chi}-\frac{\sinh^{-1}\!\sqrt{\chi}}{2\chi^{3/2}}.
  \label{etaTEanalytic}
\end{equation}
These expressions give the known contribution of the TE polarization to the 
Casimir--Polder potential~\cite{Dzyaloshinskii1961}.
As $\chi\longrightarrow\infty$, this ``efficiency'' $\eta\subTE$ converges to $1/6$,
in agreement with the strong-coupling limit (\ref{VCPTEdielectric-perfect}),
and is approximately $\chi/40$ for small $\chi$.
The remainder of the full electromagnetic Casimir--Polder potential is supplied by the
contribution from the TM polarization.  

This procedure also applies to an atom embedded in the dielectric side of a 
planar vacuum--dielectric interface.  
In this case, the explicitly renormalized form
of the potential (\ref{VCPTEdielectric}) is
\begin{align}
  V\subCP\supTE(\rA) 
  =& \frac{\hbar c \alpha_0}{4(2\pi)^{D/2}\epsilon_0}\int_0^\infty \!\!\frac{d\cT}{\cT^{1+D/2}} 
  \nonumber\\&
  \hspace{6mm}{}\times 
  \,\Bigdlangle \langle\epsr\rangle^{-3/2}-[\epsr(\rA)]^{-3/2}
\Bigdrangle_{\vect{x}(t)},
  \label{VCPTEdielectricinsiderenorm}
\end{align}
in order to properly remove the $\cT=0$ divergence.
This corresponds to subtracting the potential in the limit where
the interface is moved far away from the atom, which itself is still
embedded in the dielectric.
In evaluating this potential, the $d<0$ part of the path-averaged
expression (\ref{eq:Feynman-Kac TE one step}) applies,
and the same procedure leads to
\begin{align}
   V\subCP\supTE(d) 
   =&\,
   \frac{\hbar c\alpha_0}{8\pi^{2}\epsilon_0}
  \int_0^\infty\!\! ds\,\sqrt{s}
  \int_0^\infty\!\! d\lambda\nonumber\\
   &\hspace{-9mm}{}\times\frac{e^{-2\sqrt{2[\lambda+s(1+\chi)]}\,d}}{\sqrt{\lambda+s(1+\chi)}}
     \left(\frac{\sqrt{\lambda+s(1+\chi)}-\sqrt{\lambda+s}}  
  {\sqrt{\lambda+s(1+\chi)}+\sqrt{\lambda+s}}\right),
  \label{VCPanalyticinsideint}
\end{align}
where $d>0$ is the distance between the atom and interface.
The integration here may also be carried out analytically, so the
result may be written
\begin{equation}
   V\subCP\supTE(d) 
  =
   \frac{3\hbar c\alpha_0}{32\pi^{2}\epsilon_0z^4}\,
   \eta'\subTE(\chi),
 \label{TEfuncderivsVCPgeneraljustdielectricrepforevalgenchirepforMellinfinalD4einside}
\end{equation}
where the relative contribution compared to the (magnitude) of the total
electromagnetic strong-coupling result is
\begin{align}
 \eta'\subTE(\chi)=&\,
   \left(
      \frac{5}{6}+\frac{1}{\chi}
    -\frac{\sqrt{1+\chi}}{2\chi}
    -\frac{(1+\chi)^{3/2}}{2\chi^{3/2}} \tan^{-1}\sqrt{\chi}
   \right)\nonumber\\
   &\hspace{5mm}\times (1+\chi)^{-3/2}.
   \label{etaTEprimeanalytic}
\end{align}
Note that, on the dielectric side of the interface, the overall sign
of the potential is positive, because the efficiency factor (\ref{etaTEprimeanalytic})
is strictly positive.

\subsection{Casimir energy}

The same technique, used in evaluating the worldline
path integral (\ref{eq:Casimir worldline TEdiel}), 
yields the Casimir energy between two parallel 
dielectric interfaces.
A proper renormalization here involves subtracting the one-body contributions
 from the two-body energy, leaving only the interaction energy of the two planes.
Denoting the permittivity due to both dielectric half-spaces 
$\epsilon_{\text{r},12}(z)$, while using $\epsilon_{\text{r},1}(z)$ and 
$\epsilon_{\text{r},2}(z)$ to denote the respective single-body dielectrics,
the renormalized Casimir energy is 
\begin{align}
    E\subTE &= \frac{\hbar c}{2(2\pi)^{D/2}}\int_0^\infty \!\!\frac{d\cT}{\cT^{1+D/2}}\, \int d\vect{x}_0\nonumber\\
& \hspace{5mm}   \,\biggdlangle \bigg(\frac{1}{\sqrt{\epsilon_{\text{r},12}(\vect{x}_0)}}-\frac{1}{\sqrt{\langle\epsilon_{\text{r},12}\rangle}}\bigg) \nonumber\\
&\hspace{10mm}{}- \bigg(\frac{1}{\sqrt{\epsilon_{\text{r},1}(\vect{x}_0)}}-\frac{1}{\sqrt{\langle\epsilon_{\text{r},1}\rangle}}\bigg)\nonumber\\
&\hspace{10mm}{}-\bigg(\frac{1}{\sqrt{\epsilon_{\text{r},2}(\vect{x}_0)}}-\frac{1}{\sqrt{\langle\epsilon_{\text{r},2}\rangle}}\bigg)\biggdrangle_{\vect{x}(t)},
  \label{Casimirrenorm}
\end{align}
so that the one-body contributions are now explicitly subtracted from the two-body
energy.  The divergence at $\smash{\cT=0}$ is also removed in each case by subtracting 
the value of the integrand at $\smash{\cT=0}$, which depends on the dielectric functions
evaluated at $\x0$.
Explicitly, the permittivity functions are
$\epsilon_{\text{r},1}(z)=1+\chi_1\Theta(d_1-z)$,
$\epsilon_{\text{r},2}(z)=1+\chi_2\Theta(z-d_2)$,
and $\epsilon_{\text{r},12}(z)=1+\chi_1\Theta(d_1-z)+\chi_2\Theta(z-d_2)$
where
$\chi_1$ and $\chi_2$ are the susceptibilities of the two dielectric half-spaces,
which are separated by distance $d=d_2-d_1>0$.
The energy here is still divergent, being proportional to the transverse area of the half-spaces.
Taking the integration over the transverse dimensions to be the cross-sectional area,
$A:=\int d^{D-2}\vect{x}_0$, the energy per unit area $E\subTE/A$ produces a finite result.

The ensemble averages in Eq.~(\ref{eq:Casimir-Feynman-Kac}), integrated over $\cT$ and $x_0$,
are computed in Appendix~\ref{Appendix:TE 1body}, and the one-body and two-body
contributions are given respectively by
Eqs.~(\ref{eq:TE_1body_integral}) and (\ref{eq:TE_2body_integral}).
Combining these results with Eq.~(\ref{Casimirrenorm}), the Casimir energy
density becomes
\begin{align}
    \frac{E\subTE}{A}  =& -\!\frac{\sqrt{2\pi}\hbar c}{8\pi^2\Gamma(2)\Gamma(1/2)} 
    \int_0^\infty \!\!d\lambda\, \lambda\int_0^\infty \!\!ds\, \nonumber\\
    &\hspace{-3mm}\times \sqrt{\frac{s}{\lambda+s}}\left(\frac{r_1r_2 e^{-2\sqrt{2(\lambda+s)}d}}
    {1-r_1r_2 e^{-2\sqrt{2(\lambda+s)}d}}\right)\nonumber\\
    &\hspace{-3mm}\times\left( \sqrt{2}d + \frac{1}{\sqrt{\lambda+s(1+\chi_1)}}
      + \frac{1}{\sqrt{\lambda+s(1+\chi_2)}}\right),
    \label{ETEbyAlambdas}
\end{align}
where 
\begin{equation}
    r_i = \frac{ \sqrt{\lambda+s} - \sqrt{\lambda+s(1+\chi_i)}}{ \sqrt{\lambda+s} + \sqrt{\lambda+s(1+\chi_i)}}.
\end{equation}
This integral can be cast in a more conventional form by changing integration variables to 
$p =\sqrt{\lambda+s}/\sqrt{s}$ and $\xi =\sqrt{2s}\,d$.
Then integrating by parts with respect to $p$
results in the expression
\begin{equation}
    \frac{E\subTE}{A}=   -\frac{\hbar c \pi^2}{720 d^3}\,\gamma_{\text{TE}}(\chi_1,\chi_2),
    \label{casimirresultanalytic}
\end{equation}
where
\begin{equation}
    \gamma\subTE= -\frac{180}{\pi^4}\int_0^\infty d\xi\,\xi^{2}
    \int_1^\infty dp\,p \log\!\left[1- r_1r_2e^{-2p\xi}\right],
    \label{gammaTE}
\end{equation}
and the Fresnel reflection coefficients are
\begin{equation}
    r_i = \frac{p - \sqrt{p^2+\chi_i}}{ p + \sqrt{p^2+\chi_i}}.
\end{equation}
These expressions agree with previous calculations~\cite{Schwinger1992a,Bordag2009}.
The factor $\gamma\subTE$ gives the energy density due to the TE polarization
relative to the total electromagnetic Casimir energy
density for two perfectly conducting parallel planes.
As $\chi_1,\chi_2\longrightarrow\infty$, this efficiency factor
converges to $1/2$, reflecting the equal contributions of the TE and TM
polarizations in the perfect-conductor limit.
The Casmir force $F = -\partial_d E\subTE$ between the interfaces here also
agrees with the TE-polarization component of the Lifshitz calculation~\cite{Lifshitz1956}.  

\section{Numerical Methods}\label{sec:Numerical}

The main motivation for the development of a worldline path integral is
to enable geometry-independent numerical methods for computing Casimir energies.
To investigate the feasibility of such algorithms, we will discuss
the numerical evaluation of the
path integrals (\ref{eq:Casimir worldline TEdiel}) and 
(\ref{VCPTEdielectric}) for the TE polarization of the electromagnetic field, 
and compare the numerical solutions to
the available analytic solutions in planar geometries.
It is important to emphasize that the methods developed here
apply in \textit{any} material geometry, but the solutions correspond to
exact electromagnetic solutions in planar layered media.
In more general geometries, the solutions correspond to 
Casimir energies for scalar fields coupled to dielectric media via the
action (\ref{eq:TE_scalar_action}) with arbitrary $\epsr(\mathbf{r})$,
or to magnetic media via the
action (\ref{eq:TM_scalar_action}) with arbitrary $\mur(\mathbf{r})$.
These can be regarded as scalar approximations to a full
electromagnetic calculation.

\subsection{Path generation}

The basic ingredient for the numeric evaluation of the path integrals is the
generation of the paths themselves.
It is sufficient to consider the generation of \textit{standard} Brownian
bridges $B(t)$, or Wiener paths pinned such that $B(0)=B(1)=0$.
Numerically, the goal is to generate samples $B_k$ of a discrete representation
of the bridge
in $N$ time steps of duration $\Delta t=1/N$, such that $B_0=B_N=0$,
with the correct statistics for Wiener-path increments, $\dlangle \Delta B_k\drangle=0$
and $\dlangle \Delta B_j\Delta B_k\drangle=\delta_{jk}\Delta t$.
One intuitive approach follows from the observation that, in the continuum limit,
a Wiener process with a drift is still a Wiener process. Thus, given
a Wiener process $W(t)$, one can readily introduce a drift to force the
path to close, by setting
\begin{equation}
  B(t)= W(t)-tW(1).
\end{equation}
Then the ``gap'' $W(1)$ in the closure of the endpoint is ``pro-rated'' 
along the path.  Since the Wiener increments $\Delta W_k$  can be generated
simply by multiplying standard-normal deviates by $\sqrt{\Delta t}$, this
provides a simple method for generating the required bridges. 
At finite $N$, the statistics generated by this procedure
are only approximately correct, as the variance of each step turns
out to be $\Delta t(1-\Delta t)$.
However, it is possible to directly generate bridges with the correct
finite-$N$ statistics with only slightly more work,
by changing variables in the Gaussian path measure
in the path integral (\ref{discretepathintegralwithmeasure})
to decouple the increments.
The result corresponds to the ``v-loop'' algorithm of Gies \etal~\cite{Gies2003},
and can be compactly written as the recursion
\begin{equation}
  B_k = \sqrt{\dfrac{c_k}{N}}\,z_k + c_k B_{k-1}\quad (k = 1,\ldots,N-1),
\end{equation}
where $B_0=B_N=0$, the $z_k$ are standard-normal random deviates, and the
recursion coefficients are given by
\begin{equation}
    c_k :=\dfrac{N-k}{N-k+1} .
\end{equation}
This recursion procedure can be regarded as a discrete representation
of the well-known stochastic differential equation
\begin{equation}
   dB= -\left(\frac{B}{1-t}\right)dt + dW ,
   \label{bridgeSDE}
\end{equation}
which represents a standard Brownian bridge $B(t)$ in terms
of a Wiener process $W(t)$.
The resulting standard Brownian bridges can then be scaled and shifted according to
\begin{equation}
  x_k = x_0 + \sqrt{\cT} B_k
  \label{scaleshiftbridge}
\end{equation}
to generate paths that start at $x_0$ and return after time $\cT$, as are 
needed to evaluate the path integrals.

Numerically, the coupling to the dielectric occurs via the path average of 
$\epsr(\mathbf{r})$ along each Brownian bridge.
  This can be computed most directly as in Eq.~(\ref{eq:path_average}):
\begin{equation}
  \langle \epsr\rangle = \frac{1}{N}\sum_{k=1}^N \epsr(x_k).
  \label{discretepathavg}
\end{equation}
In evaluating the two-body Casimir energy (\ref{Casimirrenorm}), 
the explicit renormalization entails evaluating the two-body path average
 $\langle \epsilon_{\text{r},12}\rangle$, and then subtracting the one-body path averages 
 $\langle\epsilon_{\text{r},1}\rangle$ and $\langle\epsilon_{\text{r},2}\rangle$ on a pathwise basis.

\subsection{Monte-Carlo sampling}

The remaining integrals over $\cT$ and $\mathbf{x}_0$ can be performed pathwise
by scaling and shifting each Brownian bridge as in Eq.~(\ref{scaleshiftbridge}),
computing the integrals for each path via standard quadrature techniques.
In the Dirichlet-boundary limit ($\chi\longrightarrow\infty$), 
the integration over $\cT$ is particularly simple~\cite{Gies2006}:
For each path and initial position $\mathbf{x}_0$, 
the integrand ``turns on'' at some minimum time $\cT_0$ when the scaled
path first touches the surface, in which case the problem reduces to 
an integration of $\cT^{-(1+D/2)}$ over $[\cT_0, \infty)$, which
can be done analytically.
However, for a more general dielectric, it is
convenient and efficient to evaluate both integrals via Monte-Carlo
sampling, where for each path values for $\cT$ and $\mathbf{x}_0$ are drawn from 
appropriate distributions. 

The $\cT$ integration is not an obvious candidate for
Monte-Carlo sampling, because the explicit weighting factor
$\cT^{-(1+D/2)}$ in the path integrals does not yield a normalizable
distribution on $[0,\infty)$.  However due to the renormalization procedure,
which removes the $\cT=0$ divergence, for any given path
there exists a bound $\cT_0$, below which the renormalized integrand vanishes.
This bound corresponds to a range of $\cT$ where the path does not extend
far enough to cross the relevant interface.
(For more general functionals than the path average, or for a source point
$\mathbf{x}_0$ in a continuously varying background, this 
may not be exactly true.  However, one can still find a similar bound, below which
the integrand is negligibly small.)
Then, given a particular standard Brownian bridge and source point $\mathbf{x}_0$,
the value of $\cT_0$ is fixed, and $\cT$ may be drawn from the pathwise, normalized
probability density
\begin{equation}
    p(\cT;\cT_0) = \Theta(\cT-\cT_0)\,\frac{(D/2)\cT_0^{D/2}}{\cT^{1+D/2}}.
    \label{eq:T3}
\end{equation}  
Then the integrand is evaluated at the chosen value of $\cT$, and the result
must be
multiplied by the Monte-Carlo normalization factor $\smash{[(D/2)\smash{\cT_0}^{D/2}]^{-1}}$.

In sampling the spatial integral over source points $\mathbf{x}_0$, 
no explicit spatial dependence is available
as a basis for sampling, other than the geometry of the material bodies.
However, reasoning similar to that of the $\cT$ integral 
yields a serviceable distribution---note that
it is not necessary to exactly match the spatial dependence of the integrand,
but a sampling distribution that mimics the true spatial dependence reasonably
well will lead to efficient convergence of the ensemble average.
For two bodies, for example, the region between the bodies 
should contribute the most, since these paths will interact with both bodies at relatively small
values of $\cT$. Thus, their contribution
will be magnified due to the $\smash{\cT}^{-(1+D/2)}$ factor,
relative to paths associated with the exterior region.
A reasonable choice is to sample uniformly from this interior region.
Source points farther away in the exterior region should be sampled less often, because
of their smaller contribution.
A rough estimate is given by the crossing probability 
$\smash{e^{-2d_0^2/\cT}}$ [Eq.~(\ref{crossprob})]
at some distance scale $d_0$, which for example could represent the distance to the nearest
interface.  In the $\cT$ integral, this gives power-law scaling behavior:
\begin{equation}
  \int_0^\infty\!\! \frac{d\cT}{\cT^{1+D/2}}\, e^{-2d_0^2/T} = \frac{\Gamma(D/2)}{2^{D/2}\,d_0^{D}}.
\end{equation}
The power law here can then serve as a basis for the sampling distribution in the external region.
As an example, for the Casimir energy in $D=4$ dimensions between two dielectric half-spaces,
with the vacuum gap centered at the origin, the function 
\begin{equation}
   p(x_0;d_0) = \frac{3d_0^3}{8}\times\left\{ \begin{array}{lr}
        d_0^{\,-4}, & |x_0|<d_0\\
        x_0^{-4}, & |x_0|>d_0
  \end{array}
  \right.,
\end{equation}
can serve as a sampling density for $x_0$, where $d_0$ is an adjustable parameter.
For demonstration purposes, we take $d_0$ to be the distance $d$ between interfaces 
in the computations here, although the choice of $d_0=d/2$ would be more optimal
 for this problem.  
This general idea extends to higher dimensions in a general way, for example,
 by letting $d_0$ define the radius of a sphere,
which encompasses all the material objects, and 
from which samples are drawn uniformly. 
The same power-law tails then define the sampling distribution outside the sphere.
In specific geometries, better-adapted sampling densities can be used to improve
the accuracy of the calculations.

\subsection{Numerical results} 

\begin{figure}
  \includegraphics[width=\columnwidth]{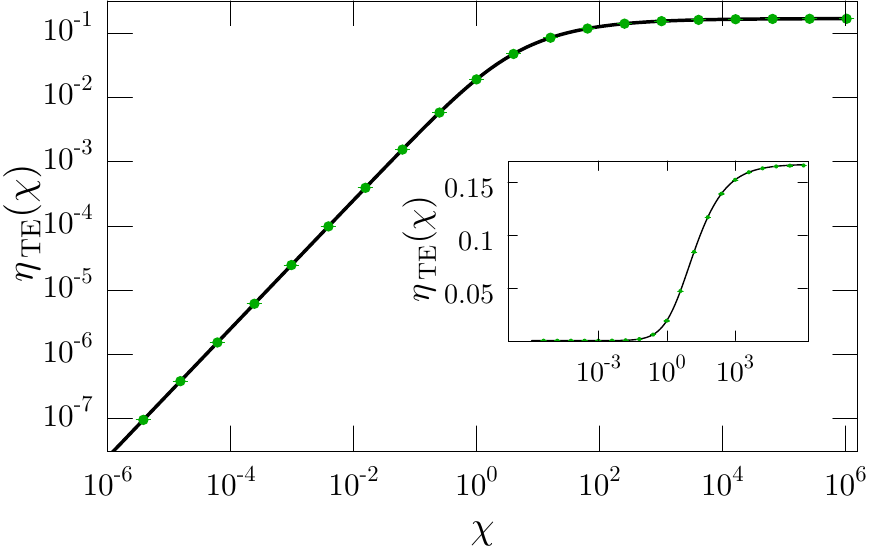}
  \caption{%
    Numerical evaluation of the Casimir--Polder path integral (\ref{VCPTEdielectricrenorm})
    for an atom near a dielectric half-space, relative to the 
    perfect-conductor Casimir--Polder potential $-3\hbar c\alpha_0/32\pi^2\epsilon_0 d^4$,
    as a function of the dielectric susceptibility $\chi$,
    for $N=10^5$ points per path, averaged over $10^8$ paths. 
    The solid line gives the analytic result (\ref{etaTEanalytic})
    for comparison.  Error bars delimit one standard deviation.
    Inset: same data plotted with a linear vertical axis.
  }
  \label{fig:eff_TE}
\end{figure}

\begin{figure}
  \includegraphics[width=\columnwidth]{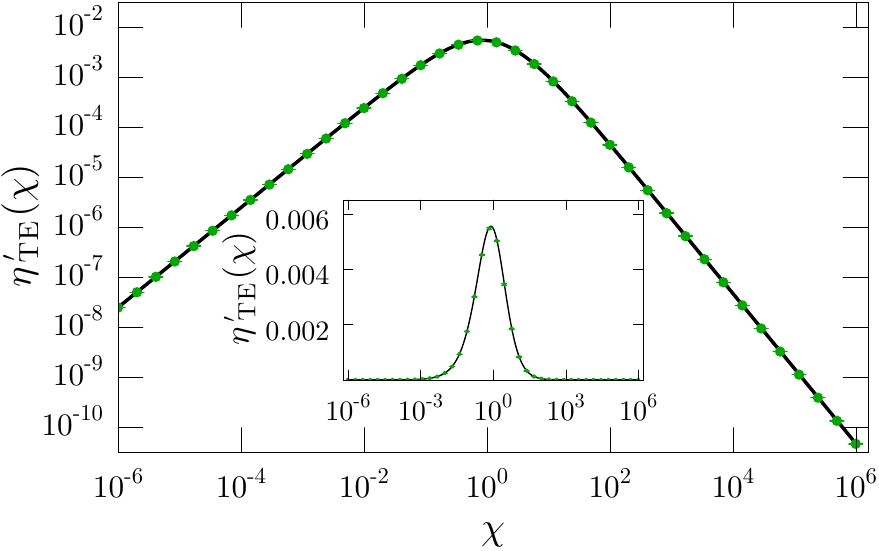}
  \caption{%
    Numerical evaluation of the Casimir--Polder path integral (\ref{VCPTEdielectricinsiderenorm})
    for an atom embedded on the dielectric side of a dielectric half-space, relative to 
    $3\hbar c\alpha_0/32\pi^2\epsilon_0 d^4$,
    as a function of the dielectric susceptibility $\chi$,
    for $N=10^5$, averaged over $10^6$ paths. 
    The solid line gives the analytic result (\ref{etaTEprimeanalytic})
    for comparison. Error bars delimit one standard deviation.
    Inset: same data plotted with a linear vertical axis.
  }
  \label{fig:eff_TE_inside}
\end{figure}

\begin{figure}
  \includegraphics[width=\columnwidth]{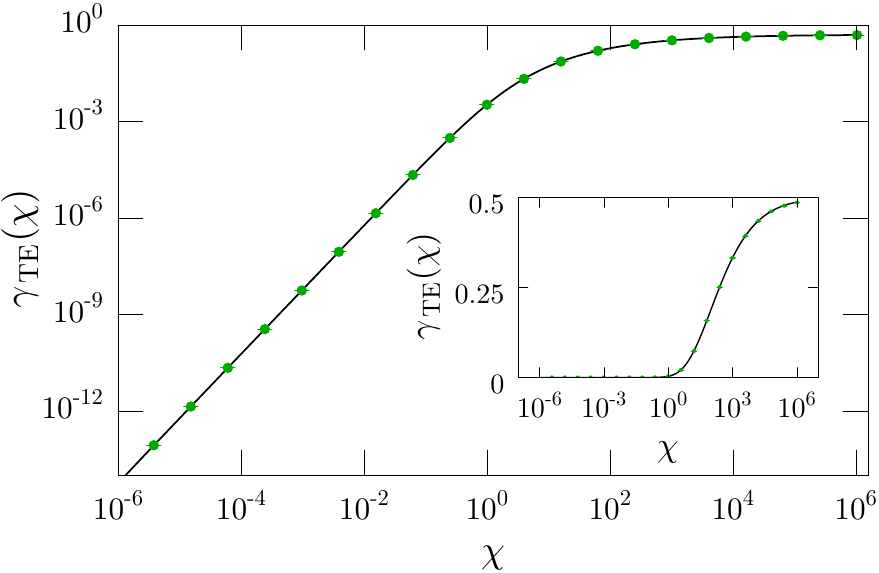}
  \caption{%
    Numerical evaluation of the Casimir-energy path integral (\ref{eq:Casimir worldline TEdiel})
    for two parallel dielectric
    half-spaces, normalized to the 
    perfect-conductor energy, as a function of the dielectric susceptibility $\chi$,
    for $N=10^5$, averaged over $10^8$ paths. The solid line gives the 
    analytic result (\ref{gammaTE})
    for comparison. Error bars delimit one standard deviation.
    Inset: same data plotted with a linear vertical axis.
  }
  \label{fig:eff_TEbody}
\end{figure}

The results for summing the 
TE-polarization Casimir--Polder path integral (\ref{VCPTEdielectricrenorm}) are shown in
Fig.~\ref{fig:eff_TE}, normalized to the total electromagnetic Casimir--Polder
potential for a perfectly conducting boundary, as in Eq.~(\ref{VCPanalytic}).
The analytic result (\ref{etaTEanalytic}) for $\eta\subTE$ is also shown for comparison;
the numerical result and analytic results agree to within a fraction of a percent.
The analogous plot 
for numerically evaluating
the Casimir--Polder path integral (\ref{VCPTEdielectricinsiderenorm}) 
for an atom embedded in the dielectric side of the interface,
compared to the analytic result (\ref{etaTEprimeanalytic}),
is shown in Fig.~\ref{fig:eff_TE_inside}, where the agreement is also excellent.
The results
for numerically evaluating the path integral (\ref{eq:Casimir worldline TEdiel})
for the normalized Casimir energy between two parallel, dielectric half-spaces
are shown in  Fig.~\ref{fig:eff_TEbody},
with the analytic result (\ref{etaTEanalytic}) for $\gamma\subTE$ for comparison;
the agreement here is similarly good.
The same set of paths were used to evaluate the path integral for each
value of $\chi$, so the data points shown are not statistically independent.
Note that the distance dependence of the path integrals follows immediately
from the dependence of the path integrals on $\cT$, so we do not explicitly test
any distance dependence here.
For finite $N$, the ensemble average tends to be biased below the true Casimir energy, particularly
for large values of $\chi$.
The main mechanism is that any given discrete path tends to overestimate the 
lower bound $\cT_0$ where the scaled path first touches an interface.

\begin{figure}
    \includegraphics[width=\columnwidth]{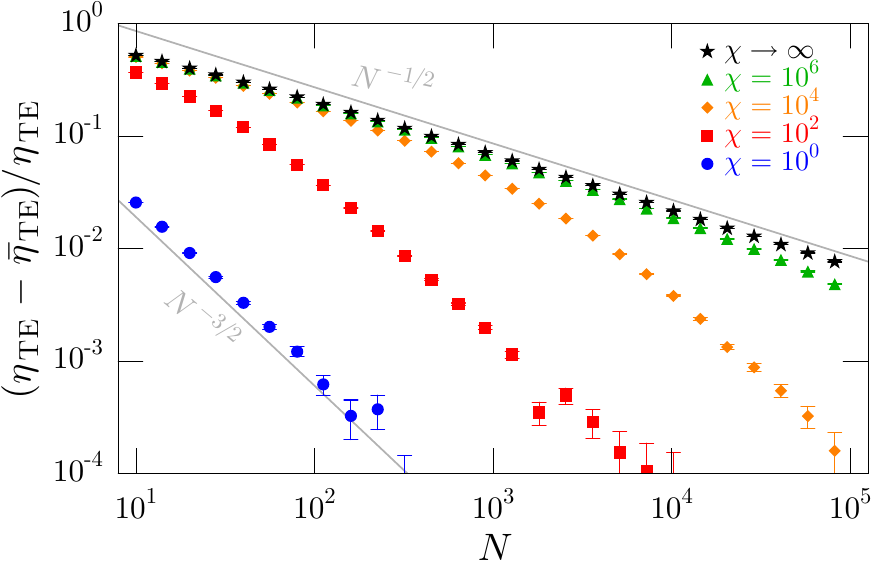}
    \caption{%
    Numerical convergence 
    of the Casimir--Polder path integral (\ref{VCPTEdielectric})
    for an atom near a dielectric half-space.
    The relative error is shown as a function of the number $N$ of points per path
    for various values of $\chi$ as indicated, including the strong-coupling
    limit $\chi\longrightarrow\infty$.
    All data points are averaged over $10^9$ paths.
    Grey lines indicated $N^{-3/2}$ and $N^{-1/2}$ scaling behaviors. 
    }
    \label{fig:conv_TEatomN}
  \end{figure}

\begin{figure}
    \includegraphics[width=\columnwidth]{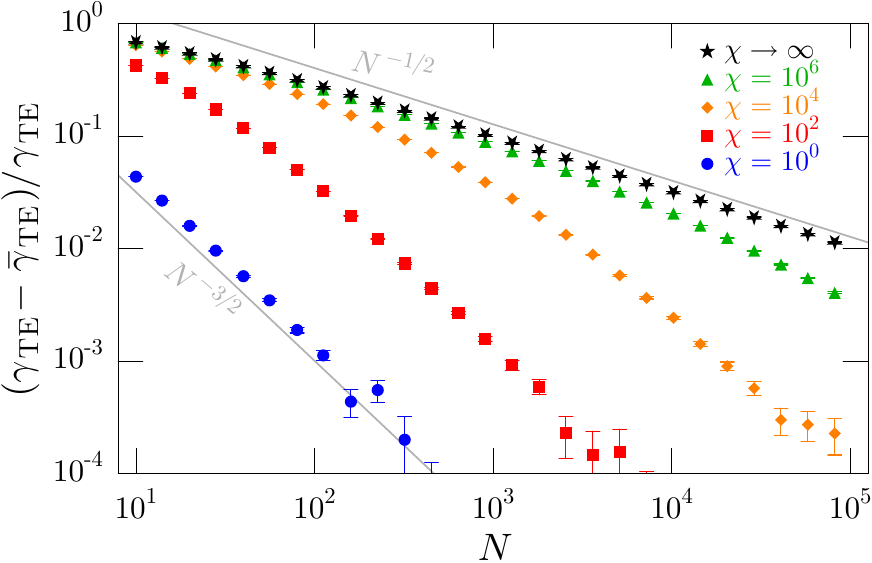}
    \caption{
    Numerical convergence of the Casimir-energy path integral (\ref{eq:Casimir worldline TEdiel})
    for two parallel dielectric half-spaces.
    The relative error is shown as a function of the number $N$ of points per path
    for various values of $\chi$ as indicated, including the strong-coupling
    limit $\chi\longrightarrow\infty$.
    All data points are averaged over $10^9$ paths.
    Grey lines indicated $N^{-3/2}$ and $N^{-1/2}$ scaling behaviors. 
    }
    \label{fig:conv_TE2wallN}
  \end{figure}

The numerical convergence of the Casimir--Polder and Casimir energies is shown in
Figs.~\ref{fig:conv_TEatomN} and \ref{fig:conv_TE2wallN}, respectively, 
where the numerical estimates $\bar{\eta}\subTE$ and $\bar{\gamma}\subTE$
approach the exact values $\eta\subTE$ and  $\gamma\subTE$ as 
the number of points $N$ per path increases.
The plots include data over a range of susceptibilities where
the error is largest: $\chi=1$, $10^2$, $10^4$, and $10^6$,
as well as the strong--coupling limit $\chi\longrightarrow\infty$.
At fixed $N$ the error increases with $\chi$, which is expected because the path
average $\langle\epsr\rangle$ can fluctuate over a wider range of values as $\chi$ increases.

The analysis of the scaling of the error with $N$ for such 
stochastic integrals is nontrivial.  Generally, we need to deal 
with two considerations: the discretization error of the integral in the 
path average (\ref{discretepathavg}), and the truncation error in
the derivation of the 
path integral in Eq.~(\ref{eq:path_int_eig}).  
Naively, one could expect that both these errors converge to zero as $N^{-1}$: 
in Eq.~(\ref{discretepathavg}) this arises from using the simplest 
trapezoidal rule, whereas in Eq.~(\ref{discretepathintegralwithmeasure}), 
there is an additional truncation error that is $O(\smash{N^{-2}})$ 
for each of the $N$ terms.  The numerical data, however, show a more 
complicated, $\chi$-dependent scaling behavior that we now discuss.

  \begin{figure}
    \includegraphics[width=\columnwidth]{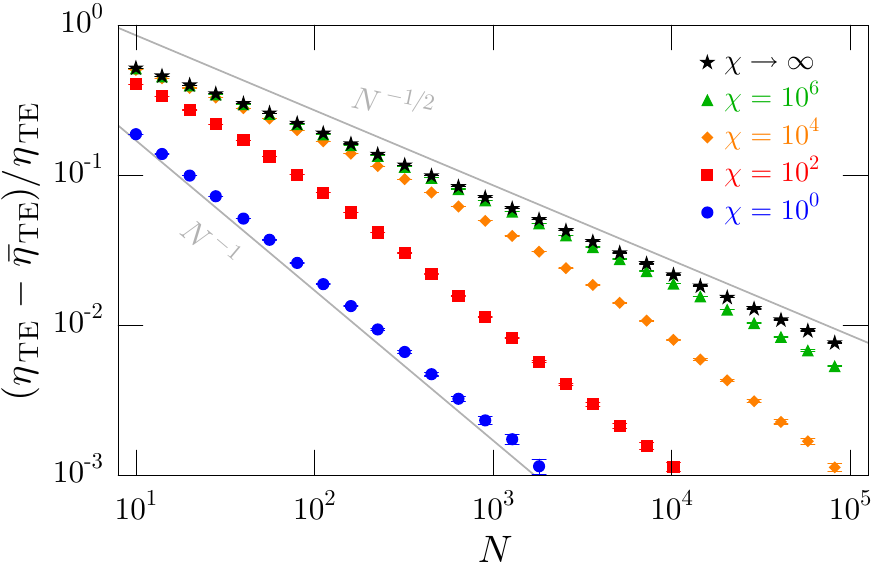}
    \caption{
    Numerical convergence of the Casimir--Polder path integral (\ref{VCPTEdielectric})
    for an atom near a dielectric half-space,
    as in Fig.~\ref{fig:conv_TEatomN}, but
    using the interpolation estimator (\ref{discretepathavgstraightlineint}) 
    instead of the trapezoidal estimator (\ref{discretepathavg}).
    All data points are averaged over $10^9$ paths.
    }
    \label{fig:conv_TEinterp}
  \end{figure}

The situation is simplest in the limit $\chi\longrightarrow\infty$, which we discuss
first.  In this limit, the error scales
as $N^{-1/2}$.
To understand this, note that the integrand of the path integral ``saturates'' whenever
a path crosses the interface.  The error made in the path integral stems
from  overestimating the first-crossing time $\cT_0$, where the time-scaled path 
just touches the surface.  
To quantify this error, we can compare the discrete path that 
falls just short of touching the interface 
with the set of continuous-time paths that pass through the same $N$ points $x_k$.  
Some of these continuous paths have more ``reach'' than the discrete
path, and so they have a non-zero probability to touch the 
surface between the discrete points.
The scaling of the error follows from considering a (discrete) path with source 
point $x_0=0$ at distance $d>0$ from the planar interface.  
The farthest extent of the path towards the interface occurs at some point $x_n$,
with $x_n\propto\smash{\sqrt{\cT}}$.
The farthest extent becomes $x_n=d$ when $\cT=\cT_0$.
Then
\begin{equation}
  \ell = d\left(1-\sqrt{\frac{\cT}{\cT_0}}\right)
\end{equation}
represents the distance between the farthest extent of the path and the interface.
For $\cT<\cT_0$, there is no contribution of the discrete
path to the path integral.
However, even for $\cT<\cT_0$, the continuous paths can still touch the interface.  
A good approximation of the touching probability is to only consider the intervals between 
$x_n$ and $x_{n\pm 1}$. The scaling behavior of the probability of both these intervals 
is the same, and for simplicity, we only consider the interval $x_n$ to $x_{n+1}$.  
Since $x_n$ is the farthest extent of the path, the probability for the continuous path
to touch the interface is maximum when $x_{n+1}=x_n$, in which case
the probability is given by the crossing probability
of a Brownian bridge for a boundary at distance $\ell$ over time $\smash{\Delta\cT=\cT/N}$.
From Eq.~(\ref{crossprob}), this is $\smash{e^{-2\ell^2/\Delta\cT}}$.
Thus, the error in the path integral is, up to an overall factor,
\begin{equation}
  e(N)=\int_0^{\cT_0}\!\!\!\frac{d\cT}{\cT^{1+D/2}}
   \,\exp\left[-\frac{2Nd^2\big(1-\sqrt{\cT/\cT_0}\big)^{\!2}}{\cT}\right].
\end{equation}
Defining $\delta\cT :=\cT_0-\cT$, the dominant contribution 
to the integral comes from
small $\delta\cT$, since large values are exponentially suppressed.
Keeping only the leading-order contribution in $\delta\cT$, changing integration
variables to $\delta\cT$, and extending the upper integration limit, 
the error becomes
\begin{align}
  e(N)\approx&\int_0^{\infty}\!\!\!\!\frac{d\delta\cT}{\cT^{1+D/2}_0}
   \,\exp\left[-\frac{2Nd^2\delta\cT^2}{4\cT_0^{\,3}}\right]\nonumber\\
  =&\,\sqrt{\frac{\pi}{2d^2\cT_0^{D-1}N}},
\end{align}
which explains the observed $N^{-1/2}$ scaling of the error in the strong-coupling limit.

To understand the scaling in the limit of small $\chi$, we will begin by considering a
similar argument.
The discrete path underestimates the value of the integrand for $\cT$ slightly
less than $\cT_0$, because while the finite-$N$ path does not cross
the interface, it may do so in the continuous limit.
The estimate for the error in this case is similar to the situation for 
the large-$\chi$ limit, but involves the sojourn time $T_\mathrm{s}$.
Again letting $x_n$ denote the point in the path with the farthest extent,
the contribution of the continuous path between neighboring
discrete points $x_n=x_{n+1}$ (which we take to be equal for the moment to give a simple 
estimate) is due to the mean sojourn time of this path segment.
In terms of the sojourn time, the error estimate is,
up to an overall factor,
\begin{equation}
  e(N)=\int_0^{\cT_0}\!\!\!\frac{d\cT}{\cT^{1+D/2}}
   \,\cT^{-1}\Bigdlangle T_\mathrm{s}\Bigdrangle,
\end{equation}
where the ensemble average here encompasses all continuous paths
connecting $x_n$ to $x_{n+1}$.
The mean sojourn time can be computed from the density (\ref{sojournbridgefinal3});
the relevant path here is a Brownian bridge
spanning time $\Delta\cT$,
with a boundary at distance $\ell\geq 0$. The result is 
\begin{equation}
  \Bigdlangle T_\mathrm{s}\Bigdrangle
   = \frac{\Delta\cT}{2}\,e^{-2\ell^2/\Delta\cT}-\sqrt{\frac{\pi \ell^2\Delta\cT}{2}}\,\,\mathrm{erfc}\,\sqrt{\frac{2 \ell^2}{\Delta\cT}}.
\end{equation}
Again expanding to lowest order in $\delta\cT :=\cT_0-\cT$ 
changing integration variables, and extending the upper integration limit to infinity,
the error becomes
\begin{equation}
  \begin{array}{l}\ds
  e(N)\approx\int_0^{\infty}\!\!\!\frac{d\delta\cT}{\cT_0^{2+D/2}}
   \Bigg[\frac{\cT_0}{2N}\,e^{-Nd^2\delta\cT^2/2\cT_0^{\,3}}
   \\\ds\hspace{26mm}{}
   -\sqrt{\frac{\pi d^2\delta\cT^2}{8N\cT_0}}\,\,\mathrm{erfc}\,\sqrt{\frac{N d^2\delta\cT^2}{2\cT_0^{\,3}}}\Bigg]
   \\\ds\hspace{8.5mm}
   =\sqrt{\frac{\pi}{32d^2\cT_0^{D-1}N^3}}.
   \end{array}
   \label{errorestimateweakchi}
\end{equation}
The resulting error estimate scales as $N^{-3/2}$, as
observed in the numerical data.
However, this argument is incomplete, as it ignores the important case when
a path segment straddles the interface, and it also ignores the contribution of the other 
path segments.

The main use of this argument is to provide a heuristic explanation of the crossover
between \textit{different} scaling behaviors for finite $\chi$ and with increasing $N$.
In the small-$\chi$ limit, 
the error associated with the first path segment touching the boundary
has an $O(N^{-1/2})$ component due to the extent of the segment,
as in the large-$\chi$ limit,
but each segment can only make an $O(N^{-1})$ contribution relative to the total path.
This extra $O(N^{-1})$ contribution  does not matter if $\chi$ is arbitrarily large, since even a single
path segment crossing the interface causes the integrand to saturate, leading to the 
$N^{-1/2}$ scaling in this regime.  For any finite $N$, there should then be a crossover
between these scaling behaviors, because it is only when $\chi/N\gg 1$ that the sojourn-time
contribution of the subpath saturates the integrand, whereas the small-chi limit
corresponds to $\chi/N\ll 1$.  Thus we expect a crossover between $N^{-1/2}$
to $N^{-3/2}$ error scaling around $N\sim\chi$, as observed
in the numerical data above.
In particular, based on the numeric results,
this means that for \textit{any} finite $\chi$, the error scales
asymptotically with the faster $N^{-3/2}$ power law after passing through the crossover regime.

In fact, the $N^{-3/2}$ error scaling is somewhat surprising:
as noted above, the discretization error of the path-average estimator
(\ref{discretepathavg}) suggests that the error should scale
no better than $N^{-1}$.
To better understand this scaling behavior, it is necessary to consider
the contribution of the entire path.  In doing so, it is  useful to
compare different approximations for the path average.
For example, an alternate estimator arises via
\begin{equation}
  \eqnarr{2.4}{
  \expct{\epsr}
    \arreq\frac{1}{\Tparam}\sum_{j=0}^{N-1}\int_{\Tparam_{j}}^{\Tparam_{j+1}}\!\!\!\!d\tau\,\epsr[x(\tau)]\\\
    \arrap\frac{1}{\Tparam}\sum_{j=0}^{N-1}\int_{x_{j}}^{x_{j+1}}\!\!\!\!dx\,\epsr(x)\,\frac{\Delta\Tparam}{\Delta x_{j}}\\\
    \arreq\frac{1}{N}\sum_{j=0}^{N-1}\frac{1}{\Delta x_{j}}\int_{x_{j}}^{x_{j+1}}\!\!\!\!dx\,\epsr(x),
  }
  \label{discretepathavgstraightlineint}
\end{equation}
where $\Delta x_j=x_{j+1}-x_{j}$ and $\Tparam_j:=j\Delta\Tparam$.
The final summand here is the average integrated
value of $\epsr(x)$ between $x_j$ and $x_{j+1}$ (in multiple spatial
dimensions, this is the average
value computed along the straight line connecting $\mathbf{x}_j$ and $\mathbf{x}_{j+1}$).
The integrals here can be computed straightfowardly for a dielectric interface in terms
of the fraction of the straight-line interval spent past the interface.
The reference method (\ref{discretepathavg}) can be written
\begin{equation}
  \expct{\epsr}
  =\frac{1}{N}\sum_{j=0}^{N-1}\frac{\epsr(x_j)+\epsr(x_{j+1})}{2}
  \label{discretepathavgreptrap}
\end{equation}
because $x_N=x_0$, and this method coincides with the ordinary trapezoidal rule
in numerical quadrature.
For a smooth, deterministic integrand, both methods have an error that scales as $N^{-2}$.
However, in computing a sojourn-time integral, the error estimate
is complicated by the involvement of a stochastic path as well as a discontinuous
integrand.   As it turns out, the interpolation
method (\ref{discretepathavgstraightlineint}) achieves the worst-case $N^{-1}$ asymptotic
scaling that we noted above.
This is shown in Fig.~\ref{fig:conv_TEinterp}, which is the same calculation
as in Fig.~\ref{fig:conv_TEatomN}, except in using the interpolation rule
(\ref{discretepathavgstraightlineint}) instead of the trapezoidal rule
(\ref{discretepathavgreptrap}).

  \begin{figure}
    \includegraphics[width=\columnwidth]{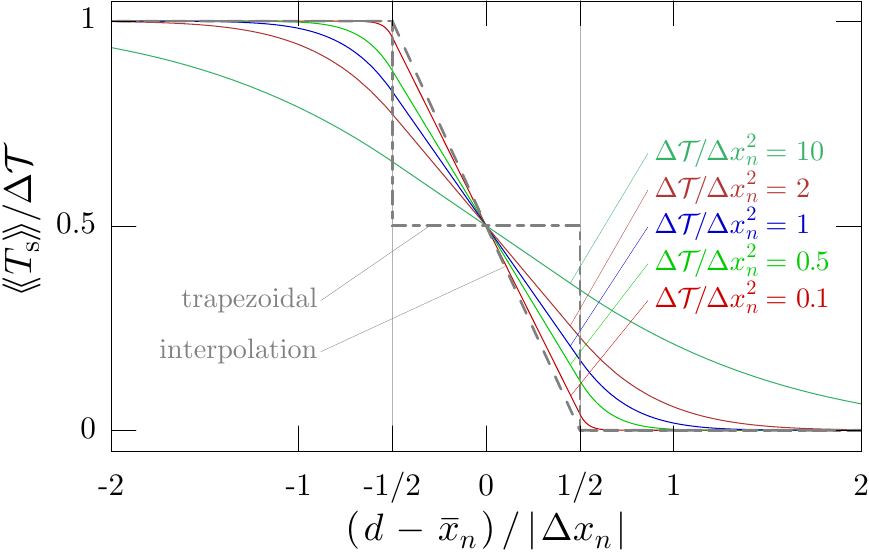}
    \caption{
    Average sojourn time $\dexpct{T_\mathrm{s}}$
    for one path segment from $x_n$ to $x_{n+1}$ [Eq.~(\ref{sojournmean})] plotted
    as a function of the interface location $d$.
    The horizontal axis is shifted such that the  interval midpoint
    $\bar{x}_n:=(x_n+x_{n+1})/2$ is centered in the plot, 
    and the axis is scaled such that the interval boundaries
    $x_n$ and $x_{n+1}$ are located at $\pm 1/2$, as marked by thin vertical lines.
    The trapezoidal and interpolation estimators for the sojourn time, given respectively by 
    Eqs.~(\ref{discretepathavgreptrap}) and (\ref{discretepathavgstraightlineint}),
    are superimposed for comparison.
    Note that the mean sojourn time decreases monotonically with $d$ 
    because the path has less opportunity
    to sojourn in the region to the right of $d$.
    }
    \label{fig:sojourn}
  \end{figure}

On the other hand, the trapezoidal rule  performs substantially better.
The difference between the 
interpolation rule and the trapezoidal rule
lies entirely in the case where the points $x_n$ and $x_{n+1}$ straddle the interface,
a case ignored by the error estimate (\ref{errorestimateweakchi}).
The mean sojourn time for a bridge between $x_n$ and $x_{n+1}$
may be calculated from the probability density
in Eq.~(\ref{sojournmean}).
The results are visualized in Fig.~\ref{fig:sojourn}.
The figure also shows the functions approximating  $\dexpct{T_\mathrm{s}}$, corresponding to the 
interpolation rule 
and the trapezoidal rule.
Note that the interpolation rule appears to be a good approximation for 
$\dexpct{T_\mathrm{s}}$ for small values of $\Delta\Tparam/\Delta x_n^{\,2}$, 
which correspond to very long (albeit rare) steps, where the Brownian path is close
to a straight-line (classical) path. The trapezoidal-rule curve does not obviously 
constitute any kind of good approximation to any of the sojourn-time curves.
The main feature to note from this plot is that, outside of the 
region between $x_n$ and $x_{n+1}$, the curves all have important common features:
they share the same reflection symmetry, and they
decay exponentially to zero or one in essentially the same way
(once the scaling in the plot is accounted for) as the case when
$x_n=x_{n+1}$.
Thus, a heuristic estimate for the error scaling follows by
adapting the error expression (\ref{errorestimateweakchi}),
which assumes $x_n=x_{n+1}$, as follows.
The estimate only accounted for the error on the leading side of the path
segment.  If we also account for the error on the trailing side, which amounts to
extending the lower integration limit to $-\infty$, the resulting expression
vanishes.  Then the leading-order result comes from keeping the
first-order term in $\delta\Tparam$ from the expansion of the $\Tparam^{-(1+D/2)}$
factor.  The resulting integral gives an error that
scales as $N^{-2}$.  However, there are $N$ total path segments, so the
overall error scales as $N^{-1}$, which matches the simple error
estimate from the $N$-point discretization of the path.
This $N^{-1}$ scaling also applies, for example, to a midpoint rule
$\epsr(\bar{x}_j)$ in place of the two-point average in the sum of Eq.~(\ref{discretepathavgreptrap}),
or to any \textit{other} estimator that depends on $\epsr(\mathbf{r})$ along the
straight line between $x_n$ and $x_{n+1}$, \textit{except} for the trapezoidal rule.

Then what is special about the trapezoidal rule?
It turns out to have the following remarkable property: when
averaged over all possible steps $\Delta x_n$, the trapezoidal rule
can exactly reproduce the mean sojourn time. Mathematically,
consider the mean sojourn time, Eq.~(\ref{sojournmean}),
with $a=-\Delta x_n/2$, $c=\Delta x_n/2$, and $t=\Delta\Tparam$,
for Brownian bridges $B_{(-\Delta x_n/2)\rightarrow (\Delta x_n/2)}(\Delta\Tparam)$ 
connecting $-\Delta x_n/2$ to $\Delta x_n/2$ in time $\Delta\Tparam$:
\begin{equation}
  \eqnarr{1.9}{\arrnone
  \overline{\bigdexpct{T_\mathrm{s}}}
  :=\intinf d\Delta x_n\,  \frac{e^{-\Delta x_n^{\,2}/2\Delta\Tparam}}{\sqrt{2\pi\Delta\Tparam}}
  \\\arrnone{}\hspace{10mm}\times
  \Bigdexpct{T_\mathrm{s}[B_{(-\Delta x_n/2)\rightarrow (\Delta x_n/2)}(\Delta\Tparam);d]}
  }
  \label{meansojournthing}
\end{equation}
The overbar here denotes the average over all possible steps
$\Delta x_n$, weighted by the Gaussian probability density for the step, 
and the step interval is centered ($\bar{x}_n=0$) to simplify the notation.
Using the estimator $\smash{\dexpct{\tilde{T}_\mathrm{s}}}$ corresponding to the trapezoidal rule,
the analogous average is
\begin{equation}
  \eqnarr{2.3}{\arrnone
  \overline{\bigdexpct{\tilde{T}_\mathrm{s}}}
  =\intinf d\Delta x_n
  \frac{e^{-\Delta x_n^{\,2}/2\Delta\Tparam}}{\sqrt{2\pi\Delta\Tparam}}
  \\\arrnone{}\hspace{10mm}\times
  \frac{\Delta\Tparam}{2}\left[\Theta\bigg(\frac{\Delta x_n}{2}-d\bigg)+\Theta\bigg(-\!\frac{\Delta x_n}{2}-d\bigg)\right]
  \\\arrnone{}\hspace{19.5mm}
  =\frac{\Delta\Tparam}{2}\,\mathrm{erfc}\!\left[\frac{\sqrt{2}\,d}{\sqrt{\Delta\Tparam}}\right],
  }
  \label{meansojournthingtrap}
\end{equation}
which ends up being exactly the same as the result of evaluating the integral in Eq.~(\ref{meansojournthing}).
When one carries out a more careful calculation of the error, analogous to
Eq.~(\ref{errorestimateweakchi}) but including the difference between
the mean sojourn time
(\ref{sojournmean}) and the trapezoidal or interpolating estimator
(i.e., with $x_n\neq x_{n+1}$ in general), 
as outlined in the previous paragraph, the results are as follows.
In the case of the trapezoidal estimator,
one finds that for any given step size $\Delta x_n$ there is a local error
of $O(N^{-2})$, which vanishes when averaged over step sizes
as in Eq.~(\ref{meansojournthing}).
When the path segment straddles the interface, the interpolating estimator
introduces an excess $O(N^{-2})$ error.  With $N$ total path segments,
this explains the $O(N^{-1})$ error scaling for the interpolating estimator,
and the higher-order error scaling for the trapezoidal estimator.

\subsection{Accelerated-convergence techniques}

The statistical error due to averaging a finite number of paths is unavoidable.
However, for a finite number of points $N$ per path, 
it is possible to use more sophisticated methods to enhance the accuracy 
relative to the performance discussed in the previous section. 
Here we will discuss two such methods in the context of the Casimir--Polder
path integral, but the same techniques also apply in the general Casimir case.
One method comes from rewriting the TE Casmir--Polder path integral (\ref{VCPTEdielectric})
in the (unrenormalized) form
\begin{align}
   &V\subCP\supTE(\mathbf{r}) 
  =
   \frac{\hbar c\alpha_0}{(2\pi)^{D/2}\sqrt{\pi}\epsilon_0}
   \intzinf \,\frac{d\Tparam}{\Tparam^{1+D/2}}\, 
  \intzinf ds\,s^2\,e^{-s^2}\,
  \nonumber\\&\hspace{30mm}\times
    \Biggdlangle
    \exp\!\Big[-\!s^2\expct{\chi}\Big]
   \Biggdrangle_{\mathbf{x}(\tau)}\!\!\!\!.
 \label{TEpathintexpform2s2repforsubpathavg}
\end{align}
This expression turns out to be the Casimir--Polder analogue of the
the Casimir free energy for dispersive media in Eq.~(\ref{CasimirFdispersion}),
if the dependence of the susceptibility on the imaginary frequency is incorporated
as $\chi(\mathbf{r})\longrightarrow\chi(\mathbf{r},is)$.
If the paths here refer to $N$-point discrete paths, 
the path average can be written
in terms of the components on each path segment as
\begin{align}
   &V\subCP\supTE(\mathbf{r}) 
  =
   \frac{\hbar c\alpha_0}{(2\pi)^{D/2}\sqrt{\pi}\epsilon_0}
   \intzinf \,\frac{d\Tparam}{\Tparam^{1+D/2}}\, 
  \intzinf ds\,s^2\,e^{-s^2}\,
  \nonumber\\&\hspace{12mm}\times
    \Biggdlangle
    \prod_{j=0}^{N-1}
    \exp\!\Bigg[-\!\frac{s^2}{\Tparam}\,\int_{\Tparam_j}^{\Tparam_{j+1}}\!\!d\tau\,\chi(\mathbf{x}(\tau))\Bigg]
   \Biggdrangle_{\mathbf{x}(\tau)}\!\!\!\!.
 \label{TEpathintexpform2s2repforsubpathavg2}
\end{align}
For a vacuum--dielectric interface, the integral in the exponential gives the sojourn time
in the dielectric of a Brownian bridge connecting $\mathbf{x}_j$ to $\mathbf{x}_{j+1}$
in time $\Delta\Tparam$.
Instead of estimating the path-segment integrals 
by using samples $\chi(\mathbf{x}_j)$ as in the trapezoidal rule (\ref{discretepathavgreptrap}),
it is most accurate
to treat the integrals in terms of the Brownian bridge between $\mathbf{x}_j$ and $\mathbf{x}_{j+1}$.
Averaging over all such bridges results in the \textit{exact}  (in the 
$N\longrightarrow\infty$ sense) expression
\begin{align}
   &V\subCP\supTE(\mathbf{r}) 
  =
   \frac{\hbar c\alpha_0}{(2\pi)^{D/2}\sqrt{\pi}\epsilon_0}
   \intzinf \,\frac{d\Tparam}{\Tparam^{1+D/2}}\, 
  \intzinf ds\,s^2\,e^{-s^2}\,
  \nonumber\\&\hspace{1mm}\times
    \Biggdlangle
    \prod_{j=0}^{N-1}
    \Bigglinklangle
    \exp\!\Bigg[-\!\frac{s^2}{\Tparam}\,\int_{\Tparam_j}^{\Tparam_{j+1}}\!\!d\tau\,\chi(\mathbf{x}(\tau))\Bigg]
    \Bigglinkrangle_{\!\!\Delta x_j}
   \Biggdrangle_{\mathbf{x}(\tau)}\!\!\!\!,
 \label{TEpathintexpform2s2repforsubpathavgavg}
\end{align}
where the connected-double-angle brackets $\smash{\linklangle\;\linkrangle_{\Delta x_j}}$ denote
the ensemble average over all bridges between $\smash{\mathbf{x}_j}$ and $\smash{\mathbf{x}_{j+1}}$.
The ensemble-averaged exponential factors here then have the form of the generating function
of the sojourn time. 
For a planar vacuum--dielectric interface, expressions for these generating functions
appear in Eqs.~(\ref{sojournmomentgen1})--(\ref{sojournmomentgen4}).
The results there are adapted to the present case by identifying $a\longrightarrow x_j$, $c\longrightarrow x_{j+1}$, 
$t\longrightarrow \Delta\Tparam$, and 
$s\longrightarrow s^2\chi/\Tparam$.
Thus, for a planar interface, a calculation performed this way has no finite-$N$ discretization
error. In fact, the analytic calculation of the TE Casimir--Polder potential in
Section~\ref{section:CPanalytic} is essentially a summation of these paths with $N=1$.
The expressions for the moment-generating functions involve integrals, but they can be evaluated
over the range of necessary values in the two free variables:
the scaled interval length $(c-a)/\sqrt{t}$ and
the scaled boundary location $(d-a)/\sqrt{t}$.  
The values needed in evaluating the path
integral can then be generated as needed 
from an interpolation table in these two variables.

The other method applies to the original path integral
(\ref{VCPTEdielectric}) for the Casimir--Polder potential,
where the goal is to accurately evaluate the path average $\,\expct{\epsr}$.
Writing out the path average as
\begin{equation}
  \expct{\epsr}
    =1+\frac{1}{\Tparam}\sum_{j=0}^{N-1}\int_{\Tparam_{j}}^{\Tparam_{j+1}}\!\!\!\!d\tau\,\chi[x(\tau)],
  \label{discretepathavgassumofint}
\end{equation}
the integrals here again have the form of the sojourn time in the dielectric medium, 
in the case of a uniform
dielectric with a sharp boundary.
The approach of Eq.~(\ref{TEpathintexpform2s2repforsubpathavgavg}),
where the sojourn-time integrals for the path segments
are replaced by the ensemble averages over Brownian bridges connecting
$\mathbf{x}_j$ to $\mathbf{x}_{j+1}$, is possible here by employing
Eq.~(\ref{sojournmean}), but not optimal.
However, since the probability density for the sojourn time is known in
Eqs.~(\ref{sojourndensity1})--(\ref{sojourndensity1}), 
these integrals can be interpreted as random variables, chosen according to the sojourn-time
probability density.
The expression for the sojourn density is relatively complicated, but
the only free parameters are the scaled interval length $(c-a)/\sqrt{t}$,
the scaled boundary location $(d-a)/\sqrt{t}$, and the sojourn time $x/t$.
It is thus feasible to compute all necessary values of the inverse cumulative probability
function, generating deviates via an interpolation table in three variables.
The same idea has been applied in financial mathematics, for example, in the pricing of 
occupation-time derivatives~\cite{Makarov2012}.
Again, for a planar interface, this method corresponds to directly taking
the limit $N\longrightarrow\infty$, with any finite-$N$ path.
The analytic calculation of the TE path 
integral described following Eq.~(\ref{sojournbridgefinal3})
via the sojourn-time distribution is equivalent to a summation
over paths in this method for $N=1$.

Of course, the planar-interface solution of the path integral is already known.
The real value of these methods lies in evaluating the path integrals
with interfaces of \textit{arbitrary} geometry.  
These methods will still
dramatically reduce the discretization error in the general case, 
provided $N$ is large enough
that the interface is well-approximated by a plane on the length scale of a path segment.
In this case, 
the planar-geometry expressions for the sojourn-time distributions can be employed.
For example, these methods are already 
approximate in the case of two parallel, planar interfaces,
because each path segment is assumed to only interact with one plane; however,
this is an excellent approximation provided that $\sqrt{\Delta\Tparam}$
is small compared to the gap between the interfaces.
These methods would be especially beneficial in the perfect-conductor 
(Dirichlet-boundary) limit,
where the asymptotic convergence with $N$ is particularly slow.

\section{Non-zero Temperature and Dispersion}
\label{sec:Thermal}
  
In considering scenarios more relevant to experiments, it is important
to incorporate material dispersion and nonzero temperatures.
Here we will discuss the generalization of the worldline formalism
to dispersive dielectric materials at nonzero temperature.
Such effects were already incorporated in the early work of 
Dzyaloshinskii \etal~\cite{Dzyaloshinskii1959,Dzyaloshinskii1961}.
However, electromagnetic quantization with dispersive materials
requires some care and has thus been the subject of much study,
because causality considerations imply that dispersive materials
are also absorptive.
Dispersive quantization is typically handled by coupling the electromagnetic
 field to an idealized, linear medium, and then coupling the medium to a bath of 
oscillators that models dissipation~\cite{Huttner1992,Dung1998,Bechler1999}.
A similar approach, outlined in Appendix~A of Ref.~\cite{Rahi2009}, emphasizes
 that the dielectric constant is related to the linear response of the underlying medium.
For a linear medium, one can carefully calculate the total energy for 
total medium--bath system, including energy lost to dissipation.  
This procedure leads to expressions for the Casimir energy that correspond to
results computed in the absence of dispersion, but with the substitution
$\epsilon\rightarrow\epsilon(i\omega)$~\cite{Barash1975,Rosa2010,Rosa2011}.

The common theme of this prior work is that the dependence of Casimir energies
 on dielectric media enters solely via the imaginary-frequency permittivity
 $\epsr(\mathbf{r},is_n)$ evaluated at the Matsubara frequencies 
$s_n:=2\pi n/\hbar\beta$.
To sketch how this comes about in the path integral,
first note that the Wick-rotated scalar field $\phi(\vect{x},\tau)$ from the
partition function (\ref{eq:field_partition_function}) can be expanded in a Fourier series as
\begin{equation}
  \phi(\vect{r},\tau) = \sum_{n=-\infty}^\infty \phi_n(\vect{r})\,e^{-is_n \tau/c} ,
\end{equation}
where
\begin{equation}
  \phi_n(\vect{r}) = \int_0^{\beta\hbar c} \!\!d\tau\, e^{is_n \tau/c} \,\phi(\vect{r},\tau).
\end{equation}
In this expression the Wick rotation has replaced the real frequency by $\omega_n\longrightarrow is_n$.
Putting this expression for $\phi(\vect{r},\tau)$ into 
the partition function (\ref{eq:field_partition_function}),
we can introduce material dispersion by giving $\epsr$ the proper
frequency dependence for each Matsubara mode.  The result with $\mur=1$ is
\begin{align}
    \ZTE =& \prod_{n=-\infty}^{\infty}\int\! D\phi_n \exp\left[ -\frac{\epsilon_0c}{2\hbar}\int d\vect{r}\right.\nonumber\\
    &\left.\times\left(\epsr(\mathbf{r},is_n)\frac{s_n^2}{c^2}|\phi_n(\vect{r})|^2 +|\nabla\phi_n(\vect{r})|^2\right)\right],
    \label{ZTE}
\end{align}
where the temperature dependence is implicit in the $s_n$.
With nonzero temperature, the appropriate thermodynamic quantity for computing forces is
the free energy, given by $\mathcal{F}=-\beta^{-1}\log Z$, which is equivalent to the
mean energy in the limit $\beta\longrightarrow\infty$.
After integration over the fields in the partition functions, the free energy becomes
\begin{equation}
  \mathcal{F} = -\beta^{-1}{\sum_n}' \log\det\left[\epsr(\mathbf{r},is_n)\frac{s_n^2}{c^2} -\nabla^2\right],  
\end{equation}
where the primed summation is defined by ${\sum'_n}f_n:=\frac{1}{2}f_0+\sum_{n=1}^\infty f_n$.
The development of the worldline path integral proceeds in the same manner as
in Section~(\ref{sec:worldline_path_integral}), with the unrenormalized result
\begin{align}
    \mathcal{F}=& -\!\frac{1}{(2\pi)^{(D-1)/2}\beta}\,{\sum_n}'\int_0^\infty \!\!\frac{d\cT}{\cT^{(D+1)/2}}\int d\vect{x}_0\nonumber\\
    &\hspace{15mm}{}\times\Bigdlangle e^{-s_n^2\langle \epsr(\vect{x},is_n)\rangle\cT /(2c^2)}\Bigdrangle_{\vect{x}(t)}.
\end{align}
The unrenormalized thermal Casimir--Polder energy follows according
 to the logic of Section~(\ref{sec:Casimir--Polder}), with the result
\begin{align}
    V\subCP(\rA,\beta) =&\, \frac{1}{2(2\pi)^{(D-1)/2}\epsilon_0c^2\beta}\,{\sum_n}'s_n^2\,\alpha(is_n)
    \nonumber\\
    &\hspace{-5mm}\times \!\int_0^\infty \!\!\frac{d\cT}{\cT^{(D-1)/2}}
    \biggdlangle  e^{-s_n^{\,2}\langle \epsr(\vect{x},is_n)\rangle\cT /(2c^2)}\biggdrangle_{\mathbf{x}(t)}\!.
  \label{VCPbeta}
\end{align}
Note that in both cases the path average $\langle\epsr\rangle$ is exponentiated,
 like a path-integral potential, in contrast to the $\langle\epsr\rangle^{-\alpha}$ forms of
 the dispersion-free path integrals.
Also, since $\epsr(\vect{r},is_n)$ is real and positive for a causal medium,
the exponential factors here are well-behaved.

In the limit of high temperature, the only contribution to the Casimir--Polder
potential comes from the lowest Matsubara mode
at frequency $s_0=0$. However, due to the presence of the factor $s_n^{\,2}$
in Eq.~(\ref{VCPbeta}), this potential vanishes.
This is consistent with known results for a planar interface \cite{Babb2004}, 
where in the limit of high temperature the 
leading-order contribution to the potential comes only from the TM polarization.

In both the Casimir and Casimir--Polder path integrals, the zero-temperature limit 
emerges as the Matsubara sum becomes well-approximated by an integral over frequency.
Making the replacement
\begin{equation}
   \frac{2\pi}{\hbar\beta}\,\,
   {\sum_n}'
   \longrightarrow \int_0^\infty \!\!ds,
\end{equation}
so that, for example, the Casimir free energy becomes
\begin{align}
    \mathcal{F}=& -\!\frac{\hbar }{(2\pi)^{(D+1)/2}}\int_0^\infty\!\! ds
    \int_0^\infty \!\!\frac{d\cT}{\cT^{(D+1)/2}}\int \!d\vect{x}_0\nonumber\\
    &\hspace{25mm}{}\times\biggdlangle e^{-s^2\cT \langle \epsr(\vect{x},is)\rangle/2c^2}\biggdrangle_{\vect{x}_0}.
    \label{CasimirFdispersion}
 \end{align}
Note that the $s$ integration in Eq.~(\ref{eq:Casimir-Feynman-Kac}), which exponentiated 
the $\langle\epsr\rangle^{-\alpha}$ dependence on the dielectric, plays essentially the
same role as the integral over the imaginary frequency $s$ here, 
but now this integration has a physical interpretation.
In the far-field limit, where the dominant transition wavelengths $\omega/c$
 are small relative to the separation of objects, the dielectric permittivity
 is given approximately by its zero-frequency value.
Then, after carrying out the integral over $s$, the path integral reduces to the 
dispersion-free expression in Eq.~(\ref{eq:Casimir worldline}) with $\mu_r\longrightarrow 1$.  

\section{Summary}

We have extended the worldline method for scalar-field Casimir energies to better model electromagnetism,
by incorporating a coupling of the field to the dielectric permittivity $\epsr(\vect{r})$
and magnetic permeability $\mur(\vect{r})$.
We have also discussed the extension of the path integrals
to dispersive media at nonzero temperature.  

The numerical evaluation of the Casimir and Casimir--Polder energies
in planar geometries, where exact results are known, 
demonstrates the good convergence properties of the path integrals.
This agreement should also extend to other geometries where
the polarizations decouple.
The numerical methods apply in arbitrary geometries, giving Casimir energies
for a scalar field coupled to a magnetodielectric material.
They also serve as a scalar approximation for the full electromagnetic Casimir energy 
in arbitrary geometries.

We have also demonstrated analytically that the worldline path integrals developed here
converge to the correct values for both Casimir and Casimir--Polder energies in planar geometries.
The analytical techniques developed here are also useful for handling the more technically challenging 
case of the TM polarization, in both analytic and numerical calculations, which we will discuss
in future work~\cite{TMpaper}.

\section{Acknowledgments}

The authors thank Richard Wagner and Wes Erickson for comments on the manuscript.
This research was supported by NSF grants PHY-1068583 and PHY-1505118.

\appendix

\section{Solutions to Feynman-Kac formulae}

In the analytic summations of worldlines in Section~\ref{sec:Analytical},
the solutions to the differential equation (\ref{eq:path_int_solution_ODE})
are required to give explicit expressions to the ensemble average 
(\ref{eq:path_int_solution}) over paths.
Here we will give an overview of the
derivation of explicit expressions that correspond to either one
or two dielectric half-spaces.
Recall that only the solution $f(x)$ at $x=0$ is required, as this is the
case that generates an average over closed paths, as required by the trace in
Eq.~(\ref{ETEtrace}).

\subsection{One-step potential}\label{Appendix:TE 1body}

The potential corresponding to a single dielectric half-space of
susceptibility $\chi$ is
\begin{equation}
  V(x) = \chi\Theta(x-d),
\end{equation}
where $d$ is the distance to the planar interface.
A path source point on the vacuum side of the interface corresponds to
$d>0$, while a source point on the dielectric side corresponds to $d<0$.
Written out explicitly, the differential equation (\ref{eq:path_int_solution_ODE}) to solve
is 
\begin{equation}
  f''(x) = 2\big[\lambda   + \chi\Theta(x-d)\big] \,f(x) -2\delta(x),
  \label{ODEonestep}
\end{equation}
and the solution gives the path average
\begin{equation}
    f(0) = \int_0^\infty\!\! dt'\,\biggdlangle \delta[W(t')] \,
    e^{-\lambda t'-\chi\int_0^{t'} \!dt''\, \Theta[x(t'')-d]} \biggdrangle,
    \label{pathavg1bdy}
\end{equation}
where again the double angle brackets $\dlangle\cdots\drangle$ denote an average over Wiener paths,
which are forced to close here by the delta function.

The solutions to Eq.~(\ref{ODEonestep}) are 
linear combinations of the functions $\smash{\exp[{\pm\sqrt{2\lambda}\,x}]}$ in regions where $x<d$,
and
of $\smash{\exp[{\pm\sqrt{2(\lambda+\chi)}\,x}]}$ in regions where $x>d$.
Then the solution is determined by enforcing continuity of $f(x)$ and $f'(x)$
across the interface at $x=d$, enforcing a jump in $f'(x)$ at $x=0$ due
to the delta function,
\begin{equation}
  f'(0_+) -f'(0_-) = -2 ,
\end{equation}
while enforcing continuity of $f(x)$ itself, and finally requiring $f(x)\longrightarrow 0$
as $x\longrightarrow\pm\infty$. With these conditions,
the solution is
\begin{equation}
    f(0) = \left\{
    \renewcommand{\arraystretch}{1.9}
     \begin{array}{ll} 
        \dfrac{1}{\sqrt{2\lambda}}\left(1+ r\, e^{-2\sqrt{2\lambda}\,d}\right)  & (d>0)\\
        \dfrac{1}{\sqrt{2(\lambda+\chi)}}\left(1 - r\, e^{2\sqrt{2(\lambda+\chi)}\,d}\right)  & (d<0),\\
      \end{array} \right. 
  \label{eq:Feynman-Kac TE 1body}
\end{equation}
where
\begin{equation}
  r = \frac{\sqrt{\lambda} -\sqrt{\lambda+\chi}}{\sqrt{\lambda} + \sqrt{\lambda+\chi}}
\end{equation}
has the form of the Fresnel reflection coefficient for TE polarization at a vacuum--dielectric
interface (provided that in terms of the angle of incidence $\theta$ from the vacuum side, one identifies
$\lambda=\cos^2\theta$). 

Combining Eqs.~(\ref{pathavg1bdy}) and (\ref{eq:Feynman-Kac TE 1body})
and applying the logic of Eqs.~(\ref{hormander1}) and (\ref{hormander2}) to remove
the delta function, the result is
\begin{align}
    &\int_0^\infty \!\!\frac{d\cT}{\sqrt{\cT}}\, e^{-\lambda \cT} \biggdlangle e^{-\chi\int_0^\cT dt\, \Theta[x(t)-d]}\biggdrangle_{x(t)} \nonumber\\
    & = \sqrt{\frac{\pi}{\lambda+\chi\Theta(-d)}}
    \left[1 + \sgn(d) \,r\, e^{-2\sqrt{2[\lambda+\chi\Theta(-d)]}\,|d|}\right],
    \label{eq:Feynman-Kac TE one step}
\end{align}
where the paths $x(t)$ are now restricted to Brownian bridges, satisfying $x(0)=x(\cT)=0$.
This solution is then useful in computing the Casimir--Polder potential for an atom near
a planar dielectric interface, by providing an expression for the $\cT$ integral
in Eq.~(\ref{eq:Casimir-Feynman-Kac}).  For example, for $d>0$ the replacements $\chi\longrightarrow s\chi$
and $\lambda\longrightarrow\lambda +s$ in Eq.~(\ref{eq:Feynman-Kac TE one step}) give
Eq.~(\ref{eq:FK TE1}).

This result is also useful in computing 
the Casimir energy of two dielectric half-spaces, where
the integral of Eq.~(\ref{eq:Feynman-Kac TE one step}) over all path source points
$x_0$ represents the one-body energy of each half-space.
Since the source point is $x_0=0$ in Eq.~(\ref{eq:Feynman-Kac TE one step}), it is
easiest to interpret $d$ as the distance from the source point to the interface,
and thus the replacement $d\longrightarrow d-x_0$ explicitly restores the source-point 
dependence,
\begin{align}
    &\int_0^\infty \!\!\frac{d\cT}{\sqrt{\cT}}\, e^{-\lambda \cT} \biggdlangle e^{-\chi\int_0^\cT dt\, \Theta[x(t)-d]}\biggdrangle_{x(t)} \nonumber\\
    & = \sqrt{\frac{\pi}{\lambda+\chi\Theta(x_0-d)}}
    \nonumber\\ & \hspace{5mm}\times{}
    \left[1 + \sgn(d-x_0) \,r\, e^{-2\sqrt{2[\lambda+\chi\Theta(x_0-d)]}\,|d|}\right], 
    \label{eq:Feynman-Kac TE one step x0}
\end{align}
where now $x(0)=x(\cT)=x_0$.
The first term on the right-hand side vanishes under renormalization, represented
by the subtractions in the last two terms in Eq.~(\ref{Casimirrenorm}),
which amounts to subtracting away 
ultraviolent divergences as $\cT\longrightarrow 0$.
The remaining integral over $x_0$ yields the one-body contribution
\begin{equation}
    I
    =  \left(\frac{1}{4\lambda}- \frac{1}{4(\lambda+\chi)}\right) r
    \label{eq:TE_1body_integral}
\end{equation}
to the total energy.
This result is useful in subtracting the one-body contributions from the two-body
interaction energy in Eq.~(\ref{Casimirrenorm}), leading to the last two terms
in the last factor in Eq.~(\ref{ETEbyAlambdas}).

\subsection{Two-step potential}
\label{Appendix:TE 2body}
 
The potential corresponding to two dielectric half-spaces with separation $d$
and
susceptibilities $\chi_1$ for $x<d_1$ and $\chi_2$ for $x>d_2$ is
\begin{equation} 
  V(x) = \chi_1\Theta(d_1-x) + \chi_2\Theta(x-d_2),
\end{equation}
where $d=d_2-d_1>0$.
Following the method in the previous section for the one-step potential,
the differential equation (\ref{eq:path_int_solution_ODE}) to solve
is 
\begin{equation}
  f''(x) = 2\big[\lambda   + \chi_1\Theta(d_1-x) + \chi_2\Theta(x-d_2)\big] \,f(x) -2\delta(x).
  \label{ODEtwostep}  
\end{equation}
Applying the same conditions as in the one-step case,  
the solution $f(0)$ may be written in three distinct regions.
For $0<d_1<d_2$, the solution corresponds to path source points in the
$\chi_1$ dielectric region $x_0<d_1$, and
is given by
\begin{align}
   f_{\text{I}}(0)&= \dfrac{1}{\sqrt{2(\lambda+\chi_1)}}  \nonumber\\
    &\hspace{1mm}\times\left[1
    +\left(\frac{r_2 e^{-2\sqrt{2\lambda}d} - r_1}{\Delta}\right)
     e^{-2\sqrt{2(\lambda+\chi_1)}\,d_1}
    \right],
\end{align}
where the reflection coefficients appear again as
\begin{equation}
    r_i := \frac{\sqrt{\lambda} -\sqrt{\lambda+\chi_i}}{\sqrt{\lambda} + \sqrt{\lambda+\chi_i}},
\end{equation}
and
\begin{equation}
    \Delta :=1-r_1r_2 e^{-2\sqrt{2\lambda}\,d}.
\end{equation}
For $d_1<0<d_2$, corresponding to path source points in the gap region $d_1<x_0<d_2$, 
the solution is given by
\begin{align}
      f_{\text{II}}(0)&= \dfrac{1}{\sqrt{2\lambda}}
      \Bigg[1+ \dfrac{2r_1r_2 e^{-2\sqrt{2\lambda}\,d}}{\Delta}\nonumber\\
      &\hspace{5mm}+ \frac{r_1 e^{2\sqrt{2\lambda}\,d_1} +r_2 e^{-2\sqrt{2\lambda}\,d_2}}{\Delta}
      \Bigg].
\end{align}
Finally, 
for $d_1<d_2<0$, corresponding to path source points in the $\chi_2$ dielectric region $d_1<d_2<0$, 
the solution is given by
\begin{align}
      f_{\text{III}}(0)&= \dfrac{1}{\sqrt{2(\lambda+\chi_2)}} \nonumber\\
      &\hspace{5mm}
      \times\left[1+ 
      \left(\frac{r_1 e^{-2\sqrt{2\lambda}d}-r_2}{\Delta}\right)
      e^{2\sqrt{2(\lambda+\chi_2)}\,d_2}
      \right].
\end{align}
In each region, the first term is independent of $d_1$ and $d_2$ and thus vanishes
under renormalization, which amounts to the subtraction
of $[\epsilon_{\text{r},12}(\vect{x}_0)]^{-1/2}$ from the path-average
functional in Eq.~(\ref{Casimirrenorm}). Again, this renormalization
corresponds to removing the
divergence at $\cT=0$ by subtracting the energy in the case where the
interfaces are moved arbitrarily far from the source point.

The Casimir energy requires the integral of this solution over all source points $x_0$,
which can again be made explicit by the replacements $d_1\longrightarrow d_1-x_0$
and $d_2\longrightarrow d_2-x_0$. Integrating the resulting expressions
in all three regions gives the total contribution
\begin{align}
    I_{12} =&\, \frac{2r_1r_2 e^{-\sqrt{2\lambda}d}d}{\sqrt{2\lambda}\Delta}
    + (r_1+r_2)\frac{(1-e^{-2\sqrt{2\lambda}d})}{4\lambda\Delta }\nonumber\\
    &+ \dfrac{r_2 e^{-2\sqrt{2\lambda}d}-r_1}{4(\lambda+\chi_1)\Delta}
    +\frac{r_1 e^{-2\sqrt{2\lambda}d} - r_2}{4(\lambda+\chi_2)\Delta}.
    \label{eq:TE_2body_integral}
\end{align}
The one-body energies 
must then
be subtracted from this result to give the total interaction
$I=I_{12}-I_1-I_2$, where 
from Eq.~(\ref{eq:TE_1body_integral}),
\begin{equation}
    I_{i} 
    =  \left(\frac{1}{4\lambda}- \frac{1}{4(\lambda+\chi_i)}\right)\,r_i
\end{equation}
for the $\chi_i$ half-space.
The result provides an expression for the last integral
in Eqs.~(\ref{eq:Casimir-Feynman-Kac}) and (\ref{Casimirrenorm}),
which yields the renormalized Casimir energy (\ref{ETEbyAlambdas}).

\section{Sojourn-time statistics}\label{appendix:sojourn}

For a stochastic process $y(t)$, the \textit{sojourn time} 
is defined as the functional
\begin{equation}
  T_\mathrm{s}[y(t);d]:=\int_0^t\!d\tau\,\Theta[y(\tau)-d],
  \label{sojourndef}
\end{equation}
where $\Theta(x)$ is the Heaviside function.
It measures the portion of the time interval $[0,t]$ that
the process spends past a boundary at position $d$.
The sojourn time is an example of the more general notion of the
\textit{occupation time} of a set, which is the 
time that a stochastic process spends within a specified set.
The sojourn time is, more specifically, the
occupation time of the set $[d,\infty)$.

In the application to path integrals in this paper, the case of interest
is when $y(t)$ has the statistics of a Wiener process \cite{Jacobs2010}.
That is, $y(t)$ corresponds to the continuous limit of a Gaussian random walk.
In each time step $dt$, the step is unbiased, $\dlangle dy(t)\drangle=0$,
where the Wiener increments are
$dy(t):=y(t+dt)-y(t)$, and the double angle brackets denote an ensemble average
over all possible steps. Further, the step variance is $\dlangle dy^2(t)\drangle=dt$
(which can also be written $dy^2(t)=dt$), and the steps are independent,
such that $\dlangle dy(t)\,dy(t')\drangle=0$ provided $t\neq t'$.
Such a Wiener process is often denoted by $W(t)$, with the convention that
$W(0)=0$, so that the probability density at time $t$ is Gaussian with zero mean
and variance $t$:
\begin{equation}
  f_{W(t)}(x) = \frac{1}{\sqrt{2\pi t}}\,e^{-x^2/2t}.
  \label{Wienert}
\end{equation}
In worldline path integrals, the paths correspond to Wiener processes whose initial
and terminal points are specified. Thus, we will use $y(\tau)$ here to denote a
stochastic process with Wiener increments, subject to the boundary conditions
$y(0)=a$ and $y(t)=c$.  The sojourn time (\ref{sojourndef}) for this process
to spend time past the boundary (i.e., the time such that $y(\tau)\geq d$)
up to total evolution time $t$,
has a probability density given explicitly by the following expressions:
\begin{widetext}
\begin{eqnarray}
   f_{T_\mathrm{s}}(x)
       \arreq
   \left[1-e^{-2(d-a)(d-c)/t}\right]\delta(x-0^+)
   +
   (2d-a-c)\sqrt{\frac{2(t-x)}{\pi t^3x}}
   \,e^{(c-a)^2/2t-(2d-a-c)^2/2(t-x)}\nonumber
   \\\arrnone{}\hspace{5mm}
   +\frac{1}{t}\left[1-\frac{(2d-a-c)^2}{t}\right]e^{-2(d-a)(d-c)/t}
   \,\mathrm{erfc}\left(\sqrt{\frac{(2d-a-c)^2x}{2t(t-x)}}\right)
   \hskip.4ex\quad(0\leq x\leq t;\; a\leq d;\;c\leq d)
  \label{sojourndensity1}
   \\
   f_{T_\mathrm{s}}(x)
  \arreq  
  \sqrt{\frac{2}{\pi}}\,\frac{(c-d)x+(d-a)(t-x)}{\sqrt{t^3x(t-x)}}
  \,e^{(c-a)^2/2t-(c-d)^2/2x-(d-a)^2/2(t-x)}
  \qquad\hskip6.6ex(0\leq x\leq t;\; a\leq d\leq c)
  \label{sojourndensity2}
  \\\arrnone{}\hspace{5mm}{}
  +\frac{1}{t}\left[1-\frac{(2d-a-c)^2}{t}\right]e^{-2(d-a)(d-c)/t}
  \,\mathrm{erfc}\left(\frac{(c-d)(t-x)+(d-a)\,x}{\sqrt{2tx(t-x)}}\right)\nonumber
  \\
   f_{T_\mathrm{s}}(x)
       \arreq
   \left[1-e^{-2(a-d)(c-d)/t}\right]\delta(t-x-0^+)
   +
   (a+c-2d)\sqrt{\frac{2x}{\pi t^3(t-x)}}
   \,e^{(a-c)^2/2t-(a+c-2d)^2/2x}\nonumber
   \\\arrnone{}\hspace{-.6mm}
   +\frac{1}{t}\left[1-\frac{(a+c-2d)^2}{t}\right]e^{-2(a-d)(c-d)/t}
   \,\mathrm{erfc}\left(\sqrt{\frac{(a+c-2d)^2(t-x)}{2tx}}\right)
   \qquad\hskip-3.8ex(0\leq x\leq t;\; d\leq c;\;d\leq a)
  \label{sojourndensity3}
   \\
   f_{T_\mathrm{s}}(x)
  \arreq  
  \sqrt{\frac{2}{\pi}}\,\frac{(a-d)x+(d-c)(t-x)}{\sqrt{t^3x(t-x)}}
  \,e^{(a-c)^2/2t-(a-d)^2/2x-(d-c)^2/2(t-x)}
  \qquad\hskip6.5ex(0\leq x\leq t;\; c\leq d\leq a)
  \label{sojourndensity4}
  \\\arrnone{}\hspace{2mm}{}
  +\frac{1}{t}\left[1-\frac{(a+c-2d)^2}{t}\right]e^{-2(a-d)(c-d)/t}
  \,\mathrm{erfc}\left(\frac{(a-d)(t-x)+(d-c)\,x}{\sqrt{2tx(t-x)}}\right).\nonumber
\end{eqnarray}
\end{widetext}
Note that the delta functions in these expressions are necessary for the 
probability densities to be normalized to unity---the delta-function coefficients
give the probability for the process to not cross the boundary at all.
Also, note that the last two expressions can be inferred from the first two
by reversing the signs of $a$, $c$, and $d$, and replacing $x$ by $t-x$,
exploiting a symmetry of the problem.
These expressions agree with those given in Ref.~\cite{Borodin2002}, though
note that the expressions here contain an explicit overall factor
of $\smash{[f_{W(t)}(c-a)]^{-1}}$ that is left implicit there.

A useful statistical average for the sojourn time is the moment-generating
function, which has the form of the Laplace transform of the probability density:
\begin{eqnarray}
  \Bigdlangle{e^{-sT_\mathrm{s}}}\Bigdrangle
  \arreq
  \int_0^t dx\,e^{-sx}\,f_{T_\mathrm{s}}(x)
    \label{sojourngenerating}
  \\ 
  \arreq
  \biggdlangle\exp\left[-s\int_0^td\tau\,\Theta[y(\tau)-d]\right]\biggdrangle.\nonumber
\end{eqnarray}
For the densities (\ref{sojourndensity1})--(\ref{sojourndensity4}), the 
corresponding moment-generating functions may be written as follows:
\begin{widetext}
\begin{eqnarray}
  \Bigdlangle{e^{-sT_\mathrm{s}}}\Bigdrangle
  \arreq
    1-e^{-2(d-a)(d-c)/t}
  \hspace{8.0cm}
  \qquad(a\leq d;\;c\leq d)
  \label{sojournmomentgen1}
  \\\arrnone{}\hspace{.5cm}
    +e^{(c-a)^2/2t}
  \frac{\sqrt{t}(2d-a-c)}{\sqrt{2\pi}\,s}\int_0^t d\tau\,\frac{1}{\sqrt{\tau^3(t-\tau)^3}}
  \,e^{-(2d-a-c)^2/2\tau}
  \left(1-e^{-s(t-\tau)}\right)\nonumber
  \\
  \Bigdlangle{e^{-sT_\mathrm{s}}}\Bigdrangle
  \arreq  e^{(c-a)^2/2t}\frac{\sqrt{ t}}{\sqrt{2\pi} s}\,\int_0^t d\tau \,\frac{(d-a)(t-\tau)[(c-d)^2-\tau]-(c-d)\tau[(d-a)^2-(t-\tau)]}{\sqrt{\tau^5(t-\tau)^5}}\nonumber
  \\\arrnone{}\hspace{6.5cm}\times
  \,e^{-(d-a)^2/2(t-\tau)
  -(c-d)^2/2\tau
  -s\tau}
  \qquad(a\leq d\leq c)
  \label{sojournmomentgen2}
  \\
  \Bigdlangle{e^{-sT_\mathrm{s}}}\Bigdrangle
  \arreq
    1-e^{-2(a-d)(c-d)/t}
  \hspace{8.0cm}
  \qquad(d\leq c;\;d\leq a)
  \label{sojournmomentgen3}
  \\\arrnone{}\hspace{.5cm}
    +e^{(a-c)^2/2t}
  \frac{\sqrt{t}(a+c-2d)}{\sqrt{2\pi}\,s}\int_0^t d\tau\,\frac{1}{\sqrt{\tau^3(t-\tau)^3}}
  \,e^{-(a+c-2d)^2/2\tau}
  \left(1-e^{-s(t-\tau)}\right)\nonumber
  \\
  \Bigdlangle{e^{-sT_\mathrm{s}}}\Bigdrangle
  \arreq  e^{(a-c)^2/2t}\frac{\sqrt{ t}}{\sqrt{2\pi} s}\,\int_0^t d\tau \,\frac{(a-d)(t-\tau)[(d-c)^2-\tau]-(d-c)\tau[(a-d)^2-(t-\tau)]}{\sqrt{\tau^5(t-\tau)^5}}\nonumber
  \\\arrnone{}\hspace{6.5cm}\times
  \,e^{-(a-d)^2/2(t-\tau)
  -(d-c)^2/2\tau
  -s\tau}
  \qquad(c\leq d\leq a).
  \label{sojournmomentgen4}
\end{eqnarray}
\end{widetext}
The expressions here match those given in Refs.~\cite{Borodin2002} and
\cite{Linetsky2001}, but again there is an explicit factor of $\smash{[f_{W(t)}(c-a)]^{-1}}$
included in the expressions here.
Finally, the mean sojourn time is given more compactly by the expression
\begin{widetext}
\begin{eqnarray}
  \Bigdexpct{T_\mathrm{s}[y(t);d]}
  \arreq
  \frac{t}{2}+\sgn(2d-a-c)\,
  \frac{t}{2}\left[e^{-2[(d-a)(d-c)\,\Theta(d-a)\,\Theta(d-c)+(a-d)(c-d)\,\Theta(a-d)\,\Theta(c-d)]/t}
  -1
  \right]
  \nonumber\\
  \arrnone\hspace{2cm}{}
  -\sqrt{\frac{\pi t}{8}}
  \,(2d-a-c)\,e^{(c-a)^2/2t}\,
  \mathrm{erfc}\left(\frac{|d-a|+|d-c|}{\sqrt{2t}}\right),
  \label{sojournmean}
\end{eqnarray}
\end{widetext}
as is consistent with differentiating the moment-generating functions 
in Eqs.~(\ref{sojournmomentgen1})--(\ref{sojournmomentgen4}) or computing the 
appropriate integral in terms of the probability density in 
Eqs.~(\ref{sojourndensity1})--(\ref{sojourndensity4}).

In deriving these expressions, the general approach is to solve 
the differential equation (\ref{ODEonestep})
to obtain a solution for the integral of the path average in Eq.~(\ref{pathavg1bdy}),
as we did in Appendix~\ref{Appendix:TE 1body}. This expression has the form of a Laplace transform
in $\lambda$, whose inverse yields the moment-generating functions
(\ref{sojournmomentgen1})--(\ref{sojournmomentgen4}).
The remaining Laplace transform in $s$ may then be inverted to 
obtain the expressions (\ref{sojourndensity1})--(\ref{sojourndensity4}) for the
probability densities.

  \bibliographystyle{aipnum4-1}

  \bibliography{worldline}

\end{document}